\documentclass[a4paper,11pt]{article}
\pdfoutput=1 % if your are submitting a pdflatex (i.e. if you have
\usepackage{jcappub} % for details on the use of the package, please
\usepackage[T1]{fontenc} % if needed
\usepackage[numbers,sort&compress]{natbib}
\usepackage{float}
\usepackage{wrapfig}
\usepackage{subcaption}
\usepackage{graphicx} %for inserting images
\usepackage{caption}

\definecolor{softgreen}{rgb}{0.0, 0.5, 0.0}
\definecolor{softred}{rgb}{0.7, 0.2, 0.2}
\definecolor{softblue}{rgb}{0.2, 0.2, 0.6}
\definecolor{lightblue}{rgb}{0.4,0.6,1.0}
\definecolor{myblue}{rgb}{0.0, 0.47, 0.75} % Define a new color

\title{\boldmath  Neural Network Reconstruction of Non-Gaussian Initial Conditions from Dark Matter Halos }

\author[a,b]{Jelte Bottema,}
\author[c,d,e,a]{Thomas Flöss,}
\author[e]{P. Daniel Meerburg}

\affiliation[a]{Kapteyn Astronomical Institute, University of Groningen, P.O.Box 800, 9700 AV Groningen, The Netherlands}
\affiliation[b]{Department of Physics, Stellenbosch University, Matieland 7602, South Africa}
\affiliation[c]{Department of Mathematics, University of Vienna, Oskar-Morgenstern-Platz 1, 1090 Vienna, Austria}
\affiliation[d]{Department of Astrophysics, University of Vienna, Türkenschanzstraße 17, 1180 Vienna, Austria}
\affiliation[e]{Van Swinderen Institute, University of Groningen, Nijenborgh 3, 9747 AG Groningen, The Netherlands}
\emailAdd{jbott@astro.rug.nl}
\abstract{We develop a machine learning approach to reconstructing the cosmological initial conditions from late-time dark matter halo number density fields in redshift space, with the goal of improving sensitivity to cosmological parameters, and in particular primordial non-Gaussianity. Using an U-Net architecture, our model achieves a cross-correlation accuracy of 44\% for scales out to $k = 0.4 \text{ h}/\text{Mpc}$ between reconstructed and true initial conditions of Quijote 1 Gpc$^3$ simulation boxes with an average halo number density of $\bar{n} = 4\times 10^{-4}$ (h/Mpc)$^{3}$ in the tracer field at $z=0$ . We demonstrate that our reconstruction is likely to be optimal for this setup and that it is highly effective at reducing redshift-space distortions. Using a Fisher analysis, we show that reconstruction improves cosmological parameter constraints derived from the power spectrum and bispectrum. By combining the power spectrum monopole, quadrupole, and bispectrum monopole up to $k_{\rm{max}} = 0.52 \text{ h}/\text{Mpc}$, our joint analysis of pre- and post-reconstructed fields from the Quijote simulation suite finds improved marginalized errors on all cosmological parameters. In particular, reconstruction improves constraints on $f_{\rm{NL}}$ by factors of 1.33, 1.88, and 1.57 for local, equilateral, and orthogonal shapes. Our findings demonstrate the effectiveness of reconstruction in decoupling modes, mitigating redshift-space distortions and maximizing information on cosmology. The results provide important insights into the amount of cosmological information that can be extracted from small scales, and can potentially be used to complement standard analysis of observational data, upon further development.}

\begin{document}
\maketitle
\flushbottom
\newpage
\section{Introduction}
\label{sec:intro}

Primordial non-Gaussianity (pnG) refers to deviations from the Gaussian distribution of the primordial density fluctuations in the early universe. PnG is crucial for understanding the early universe, particularly during the inflationary epoch \citep{achucarro2022inflation}, providing insights into the physics in this early period and allowing us to test the validity of inflationary models \citep{Primordial_Meerburg}. Until now, pnG has been most precisely constrained using the anisotropies of the CMB \citep{Ferreira:1998dk, komatsu2001acoustic, Komatsu:2003fd, Creminelli:2005hu, ade2016planck, Akrami:2019izv}. These anisotropies are linearly linked to the small density fluctuations caused by inflation, making them an excellent observable for investigating pnG. Observations by the Planck satellite have provided the most stringent constraints on pnG, which suggest close to Gaussian initial conditions \citep{Akrami:2019izv}. 

However, many models of inflation predict non-zero levels of pnG. This provides motivation to further study the subject of pnG. Although upcoming CMB experiments are expected to improve the constraints on pnG \citep{Ade:2018sbj, Abazajian:2016yjj}, the CMB has inherent limitations. These limitations stem from its two-dimensional nature and the effects of Silk diffusion damping on small scales, which effectively reduce the number of observable modes \citep{kalaja2021fundamental}. Consequently, to further our understanding of the primordial universe, it is essential to explore and develop additional probes beyond the CMB. These additional methods will be crucial for placing tighter constraints on pnG and gaining deeper insights into the early universe. The small anisotropies observed in the CMB evolve into the large-scale structure (LSS) of the universe. The three-dimensional comoving volume of the universe, spanning from the Dark Ages to the present, contains exponentially more modes than the CMB, in principle making it a powerful probe of pnG. Dark matter density fluctuations trace the primordial fluctuations from the early universe. The underlying dark matter density distribution can be studied using biased tracers such as galaxies \citep{catena2010novel, read2014local} and atomic and molecular spectral lines \citep{Pritchard:2011xb, Kovetz:2017agg}. Gravity evolves the primordial density fluctuations into the present-day LSS, a process that is highly non-linear due to the non-linear nature of gravity. Progress has been made in addressing the complex theoretical challenges related to studying the non-linear LSS \citep{Bernardeau:2001qr, Taruya:2012ut, Carrasco:2012cv, Garny:2022kbk}. Although current LSS surveys have not yet been able to place constraints on pnG that are as competitive with those provided by CMB observations \citep{Ross:2012sx, Leistedt:2014zqa, Ho:2013lda, Castorina:2019wmr, Mueller:2021wpc, cabass2022constraints, DAmico:2022gki, Cabass:2022wjy, kurita2023constraints}, the results are promising and provide ample motivation to further explore and improve LSS as a probe of pnG.

A complication that arises due to non-linearity, is the issue of mode coupling or non-Gaussian covariance \citep{Chan:2016ehg}. Non-linear gravitational evolution couples modes of different wavelengths, reducing the amount of unique information contained in each mode. Recently, it has been shown that this mode coupling in particular saturates constraints on pnG \citep{biagetti2022covariance, Coulton:2022qbc, floss2023primordial}. PnG is generally extracted from N-point correlation functions in Fourier space, in particular, two lowest order correlation functions; the power spectrum and bispectrum. Strong non-Gaussian covariance implies the transfer of information from the power spectrum and bispectrum into higher-order correlation functions and onto smaller scales. Measuring all higher-order correlation functions out to small scales is not feasible, because of the sheer growth in the number of possible wavelength configurations for higher-order correlation functions. However, by reconstructing the initial (linear) field, the dispersed information can be partially retrieved from higher-order correlation functions and smaller scales and reintegrated into the power spectrum and bispectrum at larger scales. Such a reconstruction is standard practice in the context of the BAO peak \citep{Eisenstein2007}, and is shown to lead to improved parameter constraints \citep{Wang2022, shirasaki2021constraining}. These studies primarily focused on recovering the BAO peak rather than achieving high-quality overall reconstruction of the initial conditions.

High-fidelity reconstruction at the field level can be achieved through techniques such as machine learning. Neural networks trained to emulate N-body simulations have shown promise in learning general properties of gravitational evolution \citep{Jamieson:2022}. Furthermore, it has been shown that neural networks are also able to reverse this evolution, reconstructing the initial conditions of a collapsed matter field \citep{Jindal:2023, Shallue:2023, List:2023jwo, Bayer:2022vid, Legin:2023jxc}. Our recent study has demonstrated that a neural network-based approach can reconstruct the initial dark matter density distribution from the present-day dark matter density distribution \citep{Thomas}, with an accuracy comparable to state-of-the-art iterative reconstruction algorithms \citep{Schmittfull:2017}, and once trained, operates at a fraction of the computational cost. The decoupling of different wavelength modes, due to reconstruction, reduces covariance of correlation functions. This process thus restores more unique information into the power spectrum and bispectrum of the reconstructed density field, coming from higher-order N-point correlation functions and smaller scales of the late-time density field. Using a Fisher forecast, we showed that the reconstructed power spectrum and bispectrum indeed lead to significantly improved constraints on cosmological parameters.

In reality, it is impossible to measure the dark matter distribution directly. Instead, we are limited to observing biased tracers of the underlying dark matter distribution, such as galaxies. Additionally, in previous work, the reconstruction was performed using the real-space dark matter distribution, whereas realistically, observations are obtained in redshift space. In this paper, we extend our application of neural network-based reconstruction techniques to the dark matter halo number density field in redshift space, which serves as a more realistic example of a tracer of large-scale structure. Compared to the dark matter density field, the halo field faces additional complications beyond non-Gaussian covariance. The discrete nature of dark matter halos makes them sparse tracers of the underlying density fields, introducing shot noise. Compared to the dark matter density field, where parameter constraints saturate due to non-Gaussian covariance, the saturation of parameter constraints in dark matter halo fields is primarily caused by shot noise, which dominates on smaller scales \citep{coulton2022quijote}. Although shot noise is the primary factor for parameter saturation, decoupling modes and incorporating more information from higher N-point statistics into the power spectrum and bispectrum can still be beneficial. This is supported by \citep{biagetti2022covariance}, who found a strong correlation among squeezed halo bispectrum configurations and between these configurations and the long-wavelength halo power spectrum. To approximate real observations, dark matter halo fields are translated from real space into redshift space. This introduces an additional source of non-linearity and, consequently, non-Gaussian covariance.

By applying neural network-based reconstruction to the redshifts-space $z=0$ dark matter halo fields, we will show that one can indeed reconstruct a significant part of the real-space initial conditions at the field level. Additionally, by applying our reconstruction method to the data of the Quijote simulation suite \citep{villaescusa2020quijote, coulton2022quijote} and measuring the power spectrum monopole, quadrupole, and bispectrum, we forecast improved constraints on pnG by factors of 1.33, 1.88, and 1.57 for local, equilateral, and orthogonal pnG, respectively, as well as improvements on other cosmological parameters. This improvement results from the mitigation of redshift-space distortions, the reduction of mode coupling and the decrease of parameter degeneracy. Such improvements are hard to achieve by including more non-linear modes due to parameter saturation caused by shot noise and non-Gaussian covariance.
We argue that our reconstruction method could complement observational survey data analysis, squeezing out as much information as possible from the data, by improving constraining power without necessitating costly and observationally challenging approaches such as increasing number density, expanding the observed volume to add more modes, or increasing the maximum survey redshift.

This paper is organized as follows. In Section \ref{Section pnG}, we discuss the effect of non-Gaussianity on the dark matter halo density field. We discuss the training and testing of our machine learning model and the simulations we use for training in Section \ref{chapter 5}. Section \ref{chapter 6} is dedicated to the statistical analysis, highlighting our main results and their implications for parameter constraints. We conclude in Section \ref{chapter 9}.

\section{Primordial non-Gaussianity in the dark matter halo density field}
\label{Section pnG}

\subsection{Primordial Bispectra}

For Gaussian fields, the two-point correlation function or power spectrum fully describes the field. Describing non-Gaussian fields instead requires higher-order N-point correlation functions. The lowest-order statistic revealing non-Gaussian characteristics is the three-point-correlation function or bispectrum \citep{Peebles2001}. PnG is captured by the bispectrum of the primordial potential, in Fourier space:
\begin{equation}
\label{Bispectrum}
\left<\Phi(\mathbf{k}_1)\Phi(\mathbf{k}_2)\Phi(\mathbf{k}_3)\right> = (2\pi)^3 \delta^{(3)}(\mathbf{k}_1+\mathbf{k}_2+\mathbf{k}_3)B_{\Phi}(k_1,k_2,k_3).
\end{equation}
Here $\mathbf{k}_i$ represents a 3D vector in Fourier Space, $k_i$ is the magnitude of the vector, $k_i = |\mathbf{k}_i|$. $B_{\Phi}(k_1,k_2,k_3)$ represents the bispectrum of the primordial potential $\Phi(\mathbf{k})$. Spatial homogeneity implies that the properties of space do not vary with location, which results in momentum being conserved in a system. Consequently, when considering momentum space triangles, $\mathbf{k}_1,\mathbf{k}_2,\mathbf{k}_3$, they must form a closed shape in Fourier space, which is enforced by the delta function in Equation \eqref{Bispectrum}. The bispectrum can peak in various momentum configurations, and we refer to these unique configurations as the "shape" of the bispectrum \citep{babich2004shape}. In this paper, we consider three well-known pnG templates; the local, the equilateral, and the orthogonal shape.  

%\begin{figure}[h]
 %   \centering
 %   \includegraphics[width  = 0.8\textwidth]{Pictures/Primordial non Gaussianity.drawio.png}
 %   \caption{Visualization of the power spectrum and bispectrum, highlighting the different momentum triangle configurations of the bispectrum.}
 %   \label{pnG visualized}
%\end{figure}

The so-called local shape of primordial non-Gaussianity occurs when the primordial potential takes a form:
\begin{equation}
\label{eq:localpng}
    \Phi(\mathbf{x}) = \Phi_G(\mathbf{x}) + f_{\text{NL}} \left( \Phi_G(\mathbf{x})^2 - \langle \Phi_G(\mathbf{x})^2 \rangle \right)
\end{equation}
where $\Phi_G$ is purely Gaussian, resulting in a bispectrum that peaks in the "squeezed limit", when $k_1 \ll k_2 \approx k_3$ and is given by:
\begin{equation}
\label{local}
B^{\rm{local}}_{\Phi}(k_1,k_2,k_3) = 2f_{\rm{NL}}^{\rm{local}}(P_{\Phi}(k_1)P_{\Phi}(k_2)+2\;\text{perms.}).
\end{equation}
In single-field inflationary models, such as slow-roll inflation, local pnG is necessarily suppressed \citep{maldacena2003non, creminelli2004single}, whereas large local non-Gaussianity is expected if multiple fields play a role during inflation.

The equilateral bispectrum peaks in the equilateral triangle configuration $k_1 \sim k_2 \sim k_3$. Equilateral non-Gaussianities are naturally produced in the presence of non-gravitational self-interactions of the field driving inflation  \citep{Creminelli2003, AlishahihaSilversteinTong2004, Gruzinov2005, ChenHuangKachruShiu2007, cheung2008effective, LangloisRenauxSteerTanaka2008a, langlois2008primordial, senatore2010non, RenauxPetel2011, Pajer2021}. Due to the suppression of derivatives on superhorizon scales ($k \ll aH$), the resulting bispectrum peaks when all modes have comparable wavenumbers at the time of their horizon exit. The equilateral bispectrum is given by the template:
\begin{align}
\begin{split}
B^{\rm{equil}}_{\Phi}(k_1,k_2,k_3) &= 6f_{\rm{NL}}^{\rm{equil}}(-P_{\Phi}(k_1)P_{\Phi}(k_2)+ 2\text{ perms.} - 2(P_{\Phi}(k_1)P_{\Phi}(k_2)P_{\Phi}(k_3))^{\frac{2}{3}} \\
&\quad + P_{\Phi}(k_1)^{\frac{1}{3}}P_{\Phi}(k_2)^{\frac{2}{3}}P_{\Phi}(k_3)+ 5\text{ perms.}).
\end{split}
\end{align}
The self-interactions that create the equilateral bispectrum can be captured in the Effective Field Theory (EFT) \citep{cheung2008effective}.

Finally, EFT also predicts a third shape, which is orthogonal to the equilateral bispectrum in momentum space. A template of this type is provided in Ref.~\citep{senatore2010non}:
\begin{equation}
\begin{aligned}
\label{ortho}
B^{\rm{orth}}_{\Phi}(k_1,k_2,k_3) &= 6f_{\rm{NL}}^{\rm{orth}}\left((1+p)\frac{\Delta_{123}}{k_1^3k_2^3k_3^3} - p\frac{\Gamma_{123}}{k_1^4k_2^4k_3^4}\right), \\
\Delta_{123} &= (k_T - 2k_1)(k_T - 2k_2)(k_T - 2k_3), \\
\Gamma_{123} &= \frac{2}{3}(k_1k_2 + k_2k_3 + k_3k_1) - \frac{1}{3}(k_1^2 + k_2^2 + k_3^2), \\
p &= \frac{27}{-21 + \frac{743}{7(20\pi^2 - 193)}.}
\end{aligned}
\end{equation}

\subsection{Scale-dependent bias}
\label{scale dependent bias}
Contrary to the dark matter field, where non-linear growth only leaves a small imprint of pnG in the power spectrum \cite{Coulton:2022qbc}, for biased tracers such as halos, this imprint can be much more severe. The effect of pnG on the halo power spectrum was first predicted and validated using N-body simulations with local type pnG initial conditions in Ref.~\cite{dalal2008imprints} which showed a strong scaling of $1/k^2$ on large scales in the power spectrum, for local-type non-Gaussian initial conditions. A clean physical interpretation and a model-independent prediction of this result can be obtained by applying the peak-background split \cite{slosar2008constraints, schmidt2010halo,bardeen1986statistics, kaiser1984spatial,biagetti2019hunt}. The halo number density contrast, $\delta_h(\mathbf{k}|M,z)$, is defined as:
\begin{equation}
\label{halo density contrast}
\delta_h(\mathbf{k}|M,z) = \frac{n_h(\mathbf{x}|M,z)}{\bar{n}_h(M,z)}-1,
\end{equation}
where $n_h(x|M, z)$ is the number of halos with mass $M$ at redshift $z$ and $\bar{n}_h(M, z)$ is its mean. $\delta_h(\mathbf{k}|M, z)$ represents the halo overdensity at wavenumber $\mathbf{k}$ with mass $M$ at redshift $z$. In the absence of pnG, on the largest scales, the halo overdensity is related to the matter overdensity, $\delta_m(\mathbf{k})$:% Other methods include the high peaks approximation \citep{matarrese2008effects}, and multivariate bias expansion \citep{mcdonald2008primordial} \citep{giannantonio2010structure}. \myworries{check voor al die refs}
\begin{equation}
\delta_h(\mathbf{k}|M, z) = b_1(M, z) \delta_m(\mathbf{k}, z) + \epsilon(z).
\label{overdensity halo to matter}
\end{equation}
Here $b_1 = \frac{1}{\bar{n}_h} \frac{\partial n_h}{\partial \delta_m}$ is the usual Gaussian Lagrangian bias that quantifies the response of the halo number count to large-scale matter fluctuations and $\epsilon(z)$ is shot noise due to the discrete nature of the halos. In the peak-background split approach, we separate the long- and short-wavelength modes of the density field. In this context, we examine the most well-known \textbf{local} pnG example. The separation in the Gaussian primordial fluctuations reads:
\begin{equation}
\Phi_G(\mathbf{k}) = \Phi(\mathbf{k}_\ell) + \Phi(\mathbf{k}_s),
\end{equation}
Here $\ell$ presents long modes and $s$ the short modes. Following equation \eqref{eq:localpng}, local pnG induces a coupling between long-wavelength modes $k_\ell$ and short wavelength modes $k_s$ in the primordial potential \citep{komatsu2001acoustic}:
\begin{equation}
\label{mode coupling}
\Phi(\mathbf{k}_s) \approx \Phi_{G}(\mathbf{k}_s) \left( 1 + 2f_{\text{NL}}^{\text{local}} \Phi_G(\mathbf{k}_\ell) \right).
\end{equation}
Hence in the presence of local pnGs, the dependence of the halo overdensity in Equation \eqref{overdensity halo to matter} does not only rely on the large-scale matter overdensity but also the primordial potential \citep{coulton2022quijote}:
\begin{equation}
\label{overdens first}
\delta_h(\mathbf{k}|M, z) = b_1(M, z) \delta_m(\mathbf{k}, z) + b_\Phi(M, k, z) \mathcal{M}(k, z) \Phi(\mathbf{k}) + \epsilon(z).
\end{equation}
$b_\Phi(M, k, z)$ represents the bias associated with the primordial potential $\Phi(\mathbf{k})$. Unlike $b_1(M,z)$, $b_\Phi(M, k, z)$  is a function of the wavenumber $k$, the source of the scale-dependent bias. $\mathcal{M}(k,z)$ is the matter transfer function which transfers the primordial potential to the matter overdensity. We can rewrite Equation \eqref{overdens first}:
\begin{equation}
\label{overdensity bias function second}
\delta_h(\mathbf{k}|M, z) = \left[ b_1(M, z) + b_\Phi(M, k, z) \right] \delta_m(\mathbf{k}, z) + \epsilon(z),
\end{equation}

Following \cite{schmidt2010halo}, we then obtain a relation between $b_{\Phi}$ and $b_1$ and the primordial bispectrum $B_{\Phi}(k_1,k_2,k_3)$:
\begin{equation}
b_\Phi(M, z) = \frac{2 f_{\text{NL}} \delta_c}{\mathcal{M}(k, z)} (b_1 - 1) \frac{\sigma^2_W(k)}{\sigma^2_M}.
\label{b1 to b phi}
\end{equation}
$\delta_c \approx 1.686$ is the spherical collapse threshold, $\sigma^2_M$ represents the variance of the top-hat filtered matter field, and $\sigma^2_W$ denotes the variance of the smoothed field in the presence of pnG
\begin{equation}
\sigma^2_M = \int \frac{d^3k'}{(2\pi)^3} W^2(k') P_\delta(k'), \quad \sigma^2_W = \int \frac{d^3k'}{(2\pi)^3}\frac{ W^2(k') P_{\rm{lin}}(k') B_\Phi(\mathbf{k}, \mathbf{k}', \mathbf{k}' - \mathbf{k})}{2 f_{\text{NL}} \left( P_\Phi(k) P_\Phi(k') + 2 \text{ perms.} \right)}.
\end{equation}
Here, $W_M(k)$ is the Fourier transformed top-hat filter, with radius $R^3 = \frac{3 \bar{\rho}}{4\pi}M$ set by the halo mass, and $P_{\rm{lin}}(k)$ is the linear matter power spectrum. For local pnG, $\sigma^2_W(k) \sim \text{constant}$, and on large scales where $\mathcal{M}(k, z) \sim k^2$, this implies $b_\Phi \propto 1/k^2$. For equilateral and orthogonal pnGs, $\sigma^2_W(k) \sim k^2$, resulting in $b_\Phi \propto 1$. The correction $f_{\text{NL}}^{\text{local}}$ of the power spectrum on observable scales primarily affects $b_\Phi$ ($\propto 1/k^2$) rather than higher order $b^2_\Phi$ ($\propto 1/k^4$).

%In applying the simulation-based pre-reconstruction approach, we implicitly assume perfect knowledge of the bias relation in Equation (\ref{b1 to b phi}) \citep{coulton2022quijote}. However, this relation is only valid for universal mass functions, which describes the number density of dark matter halos as a function of their mass and redshift and assumes a universal form across different scales and epochs. This approximation does not always hold for real observables, like quasars or galaxies. In Refs.~ \citep{slosar2008constraints, reid2010cosmological, barreira2020compensated, barreira2022lss} it was shown that this necessitates marginalizing over this parameter \citep{cabass2022bias, barreira2022lss}, and the prior will affect the results of this marginalization \cite{coulton2022quijote}. In this work, our objective is to demonstrate the improvement achieved through reconstruction using neural networks. As long as we maintain consistency between the pre- and post-reconstruction fields, the specific form of the bias relation is not critical to our analysis.
%Because of this, although the imprint of scale-dependent bias is a promising probe of non-zero $f_{\text{NL}}^{\text{local}}$ through the $1/k^2$ scaling, determining the precise value of $f_{\text{NL}}^{\text{local}}$ is complicated by the uncertainty on $b_\Phi$ as a whole.

\section{Reconstruction With Neural Networks}
\label{chapter 5}

While LSS has the potential to become the leading cosmological observable, it is inherently non-linear due to both gravitational interactions and astrophysical processes, challenging cosmological information extraction. This is especially true for pnGs, which are smaller than late-time induced non-Gaussianities. Mode coupling in non-linear gravitational evolution reduces unique information per mode \citep{Chan:2016ehg}, further complicating the detection of pnGs \citep{biagetti2022covariance, coulton2022quijote, floss2023primordial}. Non-linear evolution disperses the information in the initial condition to higher-order N-point statistics and smaller scales. In Ref.~\citep{Thomas} we demonstrated that neural network-based reconstruction has the potential to recover lost information, by linearizing the field, which transfers information from the non-linearly evolved modes into the recovery of the primordial modes. In this paper, we extend this method to the dark matter halo field as a realistic proxy for galaxy behaviour in LSS. We will also include redshift-space distortion effects, that are typically present in realistic tracer fields.

\begin{figure}[H]
    \centering
    \includegraphics[scale = 0.73]{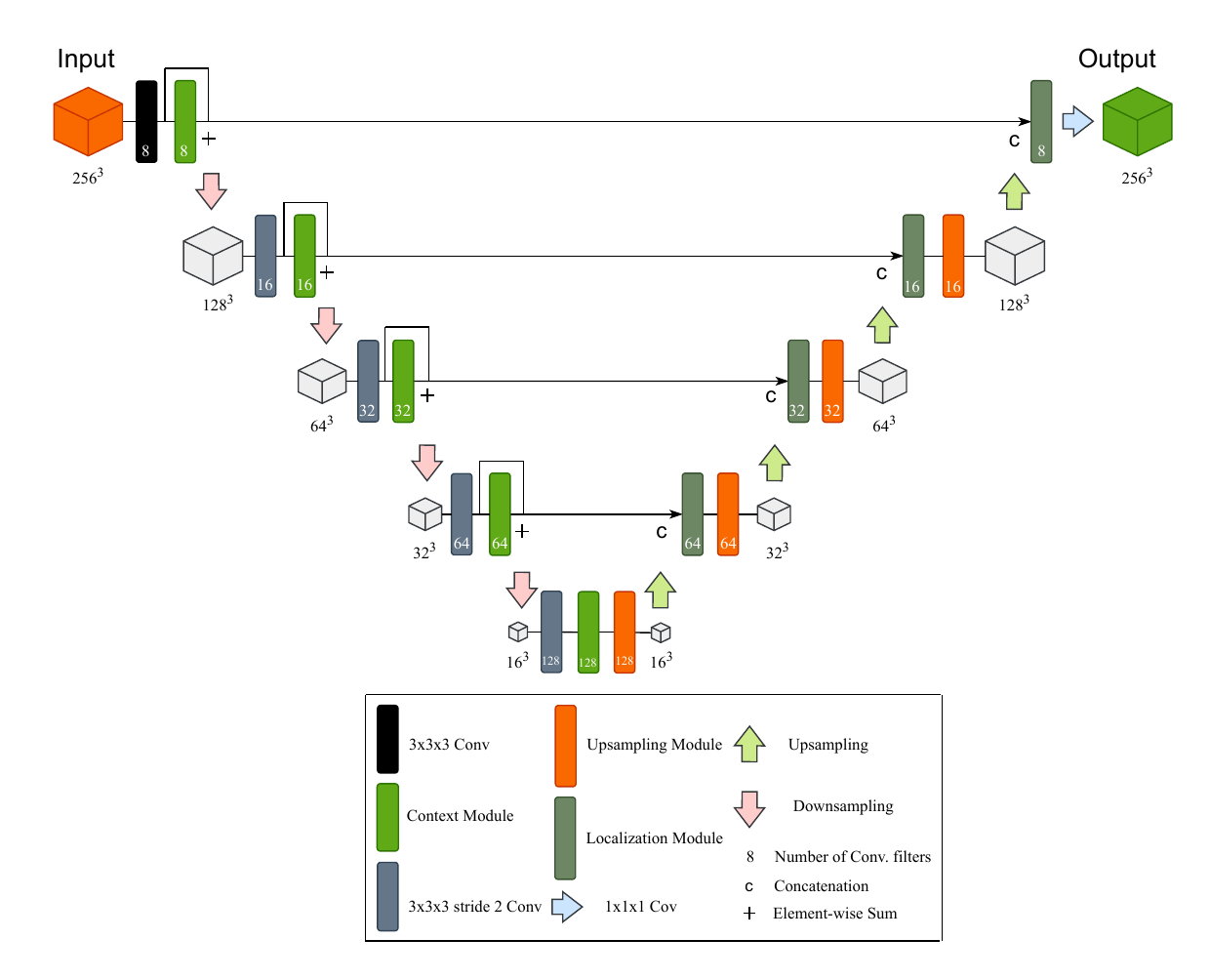}
    \caption{Diagrammatic representation of the U-Net architecture. The size of the boxes illustrates the downsampling of the density to smaller dimensions and the subsequent upsampling back to the original input dimensions.}
    \label{Unet}
\end{figure}

\subsection{U-Net Architecture}

The U-Net is a convolutional neural network originally developed for biomedical image segmentation \citep{ronneberger2015u}. U-Nets have recently achieved success in cosmology and astrophysics, especially in problems involving coupling of different length scales \citep{makinen2021deep21, gagnon2021recovering, Thomas}. Unlike ordinary convolutional neural networks (CNNs), which have a smaller receptive field, U-Nets include an encoder-decoder mechanism. This feature effectively enlarges their receptive field, enabling them to efficiently process both small-scale and large-scale information. The architecture used in this work is similar to the U-Net in previous work \citep{Thomas}, resembling the architecture originally developed in \cite{gagnon2021recovering}. The network processes an input density field on a $256^3$ grid, computing additional gradient fields on-the-fly according to grid-based perturbation theory \citep{taruya2018grid}:
\begin{equation}
\label{velocity field}
\mathbf{u}(\mathbf{x}) = \int d^3\mathbf{k} \frac{-i\mathbf{k}}{k^2} \delta_m(\mathbf{k}) e^{i\mathbf{k} \cdot \mathbf{x}},
\end{equation}
\begin{equation}
\label{deriv velocity field}
\partial_i(\mathbf{u})_j(\mathbf{x}) = \int d^3\mathbf{k} \frac{k_i k_j}{k^2} \delta_m(\mathbf{k}) e^{i\mathbf{k} \cdot \mathbf{x}},
\end{equation}
resulting in six additional input fields alongside the input density field. The U-Net architecture consists of an encoder-decoder network with skip connections. The encoder component of U-Net (left half of Figure \ref{Unet}) starts with a convolutional layer with a $3^3$ kernel, followed by a context block. This block includes two more convolutional layers, each of size $3^3$. We then combine the output of the first convolution with that of the context block, forming what is called a residual connection. Next, we introduce a stride of two in the initial convolutional layer, which downsamples the feature maps by halving their size in each spatial dimension. These downsampling operations are repeated, reducing the feature maps to a dimension of $16^3$ cells, capturing both local and global features at different scales. The decoder component of the U-Net (right half of Figure \ref{Unet}) reconstructs high-resolution feature maps from the encoded representations produced by the encoder. It utilizes upsampling operations to increase the dimensions of the feature maps by a factor of two and concatenate the output of the residual connections in the down-sampling part of the network, forming what are known as skip connections. Skip connections between encoder and decoder layers preserve spatial information and facilitate precise localization during reconstruction. At the end of the model, the U-Net has reconstructed a density field with the same dimensions as the input field $256^3$. The network is implemented using JAX and is publicly available, along with an example notebook.\footnote{\href{https://github.com/jeltebottema/dark\_matter\_halo\_reconstructions}{https://github.com/jeltebottema/dark\_matter\_halo\_reconstructions}}

\subsection{Training data}
\label{simulations section}
\subsubsection{Simulations}

For the input fields, we use the halo catalogue from the Quijote simulation suite. The initial conditions of the simulations are generated at $z=127$ using 2LPT. The N-body dark matter simulations are run with GADGET-III \citep{springel2005cosmological} as presented in \citep{quijote_matter}. The halo catalogue used in this work is constructed using the Friends of Friends (FoF) algorithm \citep{davis1985evolution} on the N-body simulations at redshift $z=0$ \citep{coulton2022quijote}. The main analysis of this work focuses on the complete set of halos available in the Quijote simulation suite, characterized by a mass greater than $M_{\rm{min}} \geq 1.32\cdot 10^{13} M_{\odot}/\text{h}$, resulting in an average number density of $\bar{n} = 4\times 10^{-4}$ (h/Mpc)$^{3}$. Additionally, the sensitivity of our reconstruction method to different number densities will be investigated. To assess the gain of doing reconstruction for cosmological parameter analysis, we will perform a Fisher forecast, using 15,000 simulations run with the fiducial cosmological model parameters, detailed in the first row of Table \ref{varying parameters}. To estimate derivatives for the Fisher forecast, we employ 500 simulations for each positive and negative perturbation of every cosmological parameter, as shown in the second row of Table \ref{varying parameters}, using finite differencing.

\begin{table}[h]
\centering
\begin{tabular}{|c|c|c|c|c|c|c|c|c|}
\hline
 & $f_{\rm{NL}}^{\rm{local}}$ & $f_{\rm{NL}}^{\rm{equil}}$& $f_{\rm{NL}}^{\rm{orth}}$ & $h$ & $n_s$ & $\Omega_m$ & $\Omega_b $ & $\sigma_8$ \\ 
\hline
$\theta_{\text{fid}}$ & 0 & 0 & 0 & 0.6711 & 0.9624 & 0.3175 & 0.049 & 0.834 \\ 
$\delta \theta$ & $\pm 100$ & $\pm 100$ & $\pm 100$ & $\pm 0.02$ & $\pm 0.02$ & $\pm 0.01$ & $\pm 0.002$ & $\pm 0.015$ \\ 
\hline
\end{tabular}
\caption{Quijote simulations' cosmological parameters. The first row presents the default cosmology applied for training the neural network and calculating the covariance matrix. The second row details the parameter variations in simulations employed for determining the derivatives.}
\label{varying parameters}
\end{table}

\subsubsection{Redshift-space Distortions}
\label{Redshift space distortions section}

In galaxy dynamics, peculiar velocities are additional galaxy motions caused by gravitational influence from nearby bodies, distinct from the Hubble expansion, leading to redshift-space distortions that complicate galaxy clustering analysis. The mapping from real-space to radial position for flat models is given by \citep{scoccimarro2004redshift}:
\begin{equation}
\label{redshift space distortions formula}
\mathbf{s} = \mathbf{x} + \frac{(1+z)}{H(z)}\mathbf{v(x)}\hat{z},
\end{equation}
where $\mathbf{v(x)}$ is the peculiar velocity, $H$ the Hubble parameter, and $\hat{z}$ is the line of sight direction. The halo catalogue provides the peculiar velocities of each halo in the $x,y,z$ axis, which is computed as a mass weighting of the halo particles \citep{coulton2022quijote}. For the initial training process, we distort the halo positions of the fiducial simulations along the $z$ axis using the peculiar velocity. For the halos in the derivative simulations, we also create the catalogues with distortions applied along either the $x$ or $y$ direction, resulting in three catalogues per realization. Although these three catalogues obviously share the same real-space configuration, they are not perfectly correlated in redshift space, particularly on small scales, providing additional realizations for estimating derivatives \citep{coulton2022quijote}. This procedure is applied only to the derivative simulations, as the 15,000 simulations used for the covariance are sufficient for our analysis.

\subsubsection{Dark Matter Halo Density Fields}
\label{making the fields}

We interpolate the halos from the halo catalogue, which covers a volume of $V = 1 \, (\text{Gpc}/\text{h})^3$, into a grid with $256^3$ voxels using the Piecewise Cubic Spline (PCS) mass-assignment scheme \citep{Chaniotis2004, sefusatti2016accurate}. The overdensity of the halos is calculated by first determining the local density of halos, $\rho_h(\mathbf{x})$, in each voxel. This local density is first subtracted and then normalized by the mean density $\bar{\rho}_h$, calculated separately for each realization of the simulations, as the mean differs throughout the different catalogues:
\begin{equation}
\label{overdensity fields formula}
\delta_h(\mathbf{x}) = \frac{\rho_h(\mathbf{x}) - \bar{\rho}_h}{\bar{\rho_h}} = \frac{\rho_h(\mathbf{x})}{\bar{\rho}_h}-1.
\end{equation}
$\delta_h$ measures the deviation of the local number density from the cosmic mean number density. In the main setup of the analysis, we omit halo mass, making $\rho_h$ a number density, $n_h$.

\subsection{Training the Model}
\label{Training}
The U-Net is trained to reconstruct the initial dark matter density fields ($z=127$), used as initial conditions for the simulations, from the final dark matter halo density fields ($z=0$) by minimizing the loss function:
\begin{equation}
    \label{Loss function}
L = \frac{1}{N_{\rm{sims}}} \sum_{i}^{N_{\rm{sims}}} \left(\frac{1}{N_{\rm{Cells}}} \sum_x (\delta_{\rm{recon}}^{(i)}(\mathbf{x})-\delta_{\rm{init}}^{(i)}(\mathbf{x}))^2 \right). 
\end{equation}

The loss function is the mean-squared error of the reconstructed $\delta_{\rm{recon}}^{(i)}$ and initial $\delta_{\rm{init}}^{(i)}$ density fields of the same snapshot of simulation $i$, calculating the difference between the density fields pixel by pixel. The dataset consists of 380 pairs of initial dark matter ($z = 127$) and final ($z = 0$) dark matter halo density fields, of which we used 250 for training and 130 for validation. Our primary goal is to determine the information content of cosmological parameters from the power spectrum and bispectrum of the reconstructed density field. The FFT-based estimator used to compute the bispectrum is limited to a scale of $k_{\text{max}} = \frac{2}{3} k_{\text{Nyq}} \approx 0.53 , \text{h}/\text{Mpc}$ \citep{sefusatti2016accurate}, where $k_{\text{Nyq}} \approx 0.82 , \text{h}/\text{Mpc}$ is the Nyquist frequency. The Nyquist frequency is given by $k_{\text{Nyq}} = \pi \frac{N_G}{L}$, which is the highest frequency that can be accurately resolved, determined by the number of grid points $N_G$ and box size $L$.  Since our focus will be on a specific subset of modes in the density fields, we only consider initial dark matter density fields up to this maximum wave number $ k_{\text{max}}$, applying an isotropic sharp cut-off in Fourier space \citep{Thomas}. This method ensures that the network avoids reconstructing modes that we will not use in our final analysis, thereby accelerating the training process. Note that the halo fields are not filtered, and thus contain information all the way up to $\sqrt{3}k_{\text{Nyq}} = 1.4 $ h/Mpc. To ensure model stability, we employ exponential moving average (EMA) of weights \citep{tarvainen2017mean}. We train the network using an Nvidia A100 40GB GPU until the validation loss stops improving significantly, which takes approximately four hours. The training is halted when overfitting starts to occur, ensuring that the model generalizes well to unseen data.

\begin{figure}[h]
    \centering
    \includegraphics[width = \textwidth]{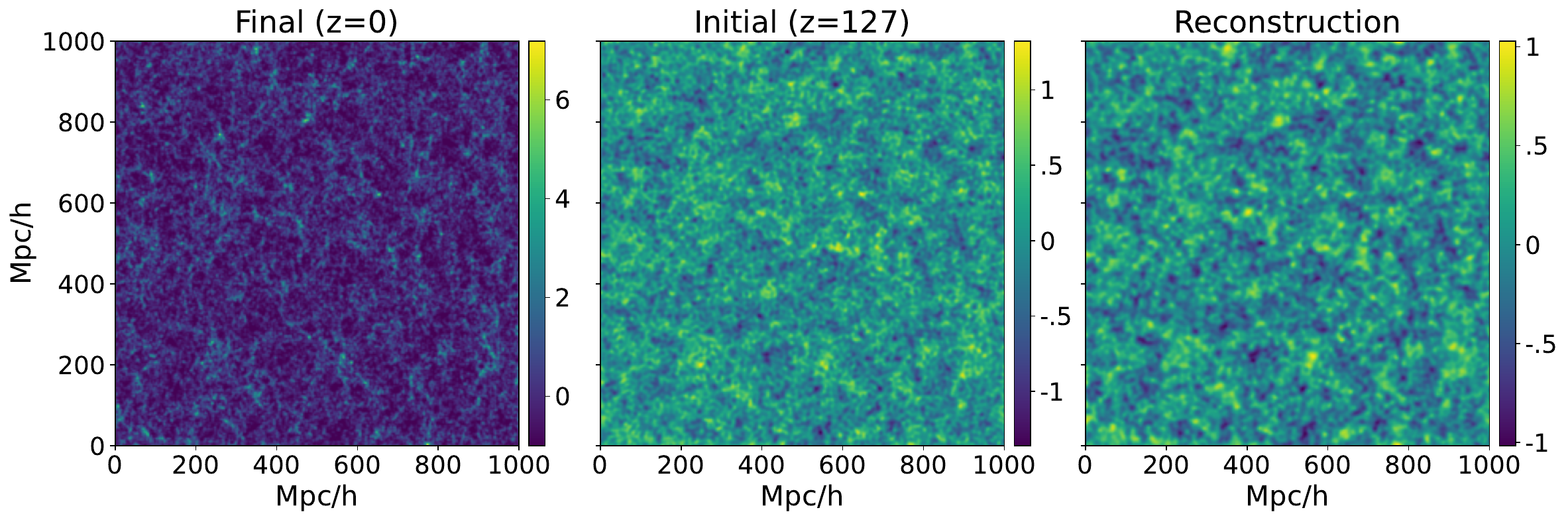}
    \caption{Left: The dark matter halo field ($z=0$). Middle: the initial dark matter field ($z=127$). Right: the reconstructed density field. The fields are three-dimensional but are averaged over along a segment of the same spatial axis for visualization purposes.}
    \label{triple fields plot}
\end{figure}

\subsection{Reconstruction Performance}

\begin{figure}[h] 
    \centering
    \includegraphics[scale = 0.6]{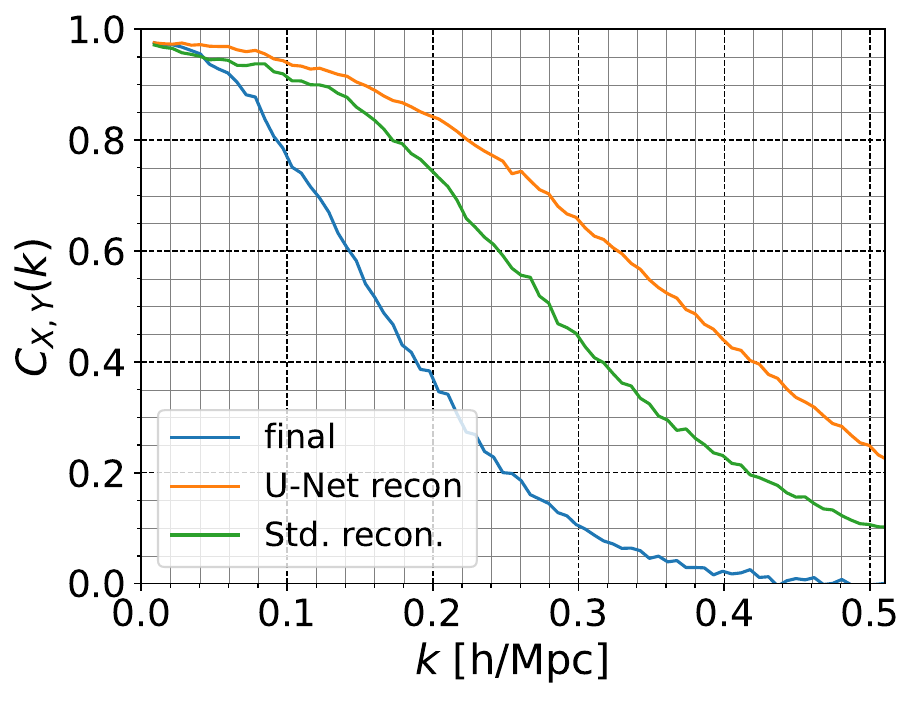}
    \caption{The cross-correlation of density fields at a single realization. The blue line represents the cross-correlation between the final dark matter halo fields ($z = 0$) and initial dark matter fields ($z = 127$), while the orange line represents the correlation between the reconstructed and initial fields. The green line represents the cross-correlation between a standard reconstructed field and the initial field}
    \label{correlation of reconstruction}
\end{figure}

The output of the model is a reconstructed density field derived from the final dark matter halo field at $z=0$. This field can be compared with the target initial dark matter field at $z=127$, an example of which is provided in Figure \ref{triple fields plot}. As expected, the reconstructed linear matter density field is visibly more linear and less non-Gaussian, and closely resembles the true initial conditions on large scales. This Figure also illustrates that the reconstruction is less accurate on smaller scales. To make these observations more quantitative, we calculate the cross-correlation between two fields $X$ and $Y$ as:
\begin{equation}
\label{covariance}
C_{X,Y}(\mathbf{k}) = \frac{P_{X,Y}(\mathbf{k})}{\sqrt{P_X(\mathbf{k}) P_Y(\mathbf{k})}}.
\end{equation}
Here $P_{X,Y}(\mathbf{k})$ is the cross-power spectrum of fields $X$ and $Y$, and $P_X(\mathbf{k})$ the auto-power spectrum of field $X$. Here, $X,Y$ can be the reconstructed, initial ($z = 127$) matter density fields or the final ($z=0$) halo density field. A perfect reconstruction thus implies $C_{\text{recon,init}}(\mathbf{k}) = 1$. Figure \ref{correlation of reconstruction} shows the cross-correlation of the final and reconstructed fields with the initial density field for a single realization, demonstrating a clear improvement in the cross-correlation after reconstruction. We find an accuracy of $44\%$ and upward on scales below $k=0.4$ h/Mpc, a significant improvement over the $3\%$ correlation of the final halo field with the initial dark matter field and on this scale, and the 23\% correlation achieved using the standard reconstruction method \cite{Eisenstein2007}.\footnote{We perform a single iteration of reconstruction using the RecIso scheme for removing redshift-space distortions as described e.g. in \cite{Chen:2024eri}} We have also trained the model without the gradient velocity fields in equations \eqref{velocity field} and \eqref{deriv velocity field}, and is shown in the Appendix Figure \ref{without PT}. Interestingly, the model shows similar performance when provided a halo field with an isotropic sharp cut-off at the  Nyquist frequency (instead of the natural cut-off of the box $\sqrt{3}k_{Nyq}$), i.e. disregarding the modes along the spatial diagonal that arise due to the three-dimensional box structure of the density fields. In other words, the reconstruction up to $k_{\text{max}} = 0.52$ Mpc/h is already saturated completely by the halo-field information up to the Nyquist frequency. This strongly suggests that our model has performed the best possible reconstruction for the given setup. At higher number densities (lower shot noise), these smaller scales could potentially provide additional information for reconstruction. 

\begin{figure}[H]
    \centering
    \begin{subfigure}[b]{0.48\textwidth}
        \centering
        \includegraphics[width=\textwidth]{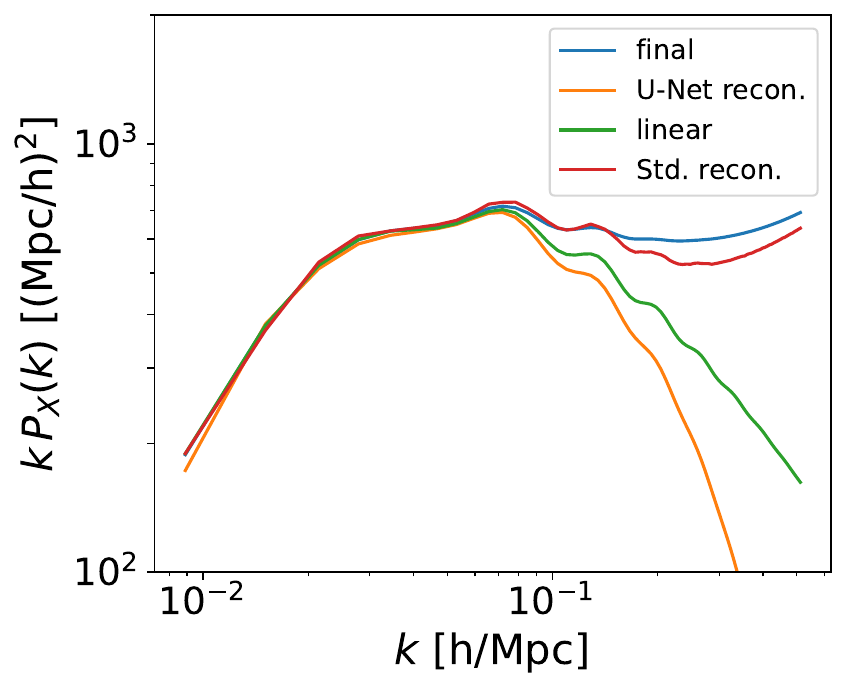}
        \caption{}
        \label{Power spectrum monopole}
    \end{subfigure}
    \hfill
    \begin{subfigure}[b]{0.48\textwidth}
        \centering
        \includegraphics[width=\textwidth]{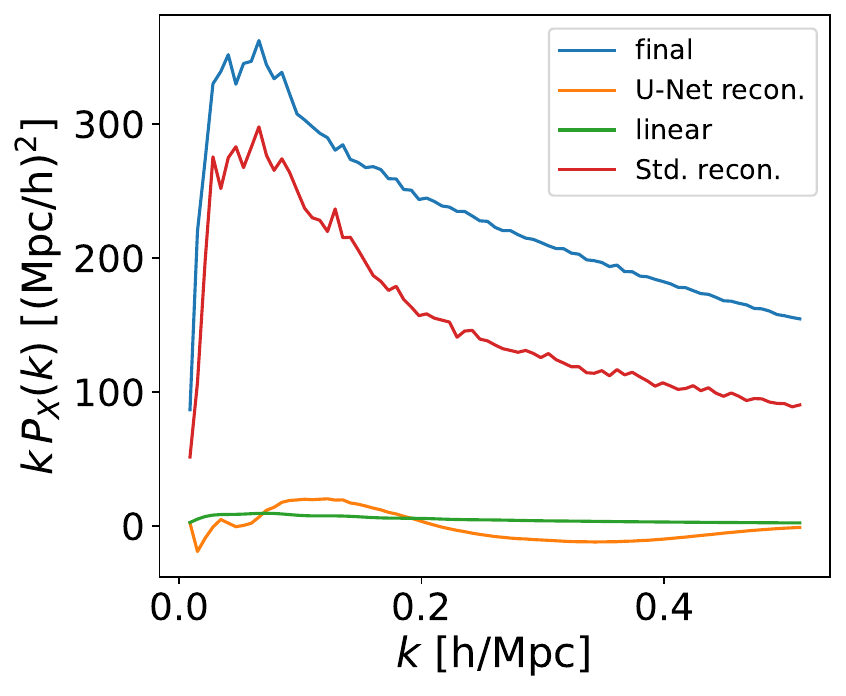}
        \caption{}
        \label{Power spectrum quadrupole}
    \end{subfigure}
    \caption{The mean power spectrum taken from 15000 fiducial cosmology simulations for the pre- and post-reconstructed density fields, compared with the initial linear power spectrum: (a) Power spectrum monopole (b) Power spectrum quadrupole. In the plot, `final' denotes the pre-reconstructed halo number density field, while 'U-Net recon' denotes the reconstructed linear matter density fields. In these plots, the reconstructed and linear fields at $z=127$ are rescaled to $z=0$, also the final and standard reconstruction spectra are corrected for the linear bias, $b_1$ (for standard reconstruction the spectra are only averaged over 50 realizations).}
    \label{Spectras compared}
\end{figure}

Redshift-space distortions introduce additional non-linearity into the halo fields. We can investigate the ability of our model to mitigate these distortions by looking at the power spectrum after reconstruction. The power spectra over many realizations are given in Figure \ref{Power spectrum monopole} and Figure \ref{Power spectrum quadrupole}, for the monopole and quadrupole, respectively. In these figures, the spectra for the reconstructed fields and the linear fields at $z=127$ are rescaled to $z=0$, and the final halo field is corrected for the linear bias term, $b_1$. In this way, the improvement in the spectra shown in these plots can be entirely attributed to our reconstruction method. We note that the quadrupole is largely removed by the reconstruction, demonstrating a strong reduction of redshift-space distortions. This level of reduction is not achieved with standard reconstruction, showcasing the advantage of our method in eliminating redshift-space distortions.

Since the amount of information in the dark matter halo field is dependent on the shot noise, which dominates on small scales \citep{coulton2022quijote}, it is worth exploring how varying levels of shot noise in the input fields affect the accuracy of reconstruction. To address the impact of shot noise on the reconstruction quality, we explored two modifications to the input halo density fields in Appendix \ref{modifying shot noise}. First, we test the effects of increasing the minimum halo mass threshold (decreasing the number density), which inherently raises shot noise. On large scales up
to $k = 0.1 \text{h}/\text{Mpc}$, the reconstruction quality is approximately the same for different number
densities. However, for lower number densities, the reconstruction clearly deteriorates more
rapidly at smaller scales. Second, we apply binned halo mass weighting to effectively reduce shot noise. We found that incorporating additional mass information
into the input density fields improves the quality of reconstruction.

\section{Cosmological Information Content of the Reconstructed Field}
\label{chapter 6}
Using the reconstruction method presented in the previous section, we now turn to analysing the information content on cosmological parameters, in order to assess the effectiveness of our reconstruction method for cosmological parameter inference. First, we measure the power spectrum monopole and quadrupole, as well as the bispectrum monopole from the reconstructed, initial, and final density fields. Then we perform a Fisher analysis on the estimated spectra, utilizing a Fisher estimator first proposed in \citep{coulton2023estimate}, which combines the standard Fisher estimator with a compressed Fisher estimator, to ensure that we obtain an unbiased forecast for the analysis.

\subsection{Power Spectrum and Bispectrum Estimator}

The redshift-space halo power spectrum $P_{hh}(k, \mu)$ depends both on the magnitude of the wave vector and the angle to the line of sight, $ \mu = \mathbf{k} \cdot \mathbf{n}$, and can be expanded into multipoles:
\begin{equation}
P_{hh}^{(\ell)}(k) = \frac{2\ell + 1}{2} \int d\mu \, \mathcal{L}_{\ell}(\mu) P_{hh}(k, \mu),
\end{equation}
where $\mathcal{L}_{\ell}(\mu)$ are the Legendre polynomials, such that the quadrupole $P^{(2)}_{hh}(k)$ depends on the line-of-sight angle $\mu$, unlike the monopole $P^{(0)}_{hh}(k)$ . We define the estimators for the power spectrum monopole and quadrupole of a density field $\delta$ as:
\begin{equation}
\label{power spectrum estimator with legendre}
\hat{P}^{(\ell)}(k_i) = \frac{\ell+1}{2N_i} \sum_{\mathbf{q}\in k_i} \mathcal{L}(\mu)\delta(\mathbf{q})\delta^{*}(\mathbf{q}),
\end{equation}
where we sum over all momenta $\mathbf{q}$ that are within a shell $[k_i-\Delta k/2+\Delta k/2]$. $N_i$ is a normalization factor that counts the number of modes that fall in the bin. For the power spectrum calculations, the largest scale bin starts at $k_f$, where $k_f$ is the fundamental frequency given by $k_f = \frac{2 \pi}{L}$, where $L$ is the physical size of the simulation box (in our case $1000$ Mpc). We use bins with a width of $\Delta k = k_f$, up to the smallest scale $k_{\rm{max}} = 82.5k_f \approx 0.52 , \text{h}/\text{Mpc}$.

Turning to the bispectrum, similar to the halo power spectrum, redshift-space distortions cause the halo bispectrum to depend on the orientation of the modes relative to the line of sight. In this work, however, we focus exclusively on the bispectrum monopole, which is defined by averaging over all possible angles and is estimated as:
\begin{equation}
\label{Bispectrum estimator}
\hat{B}(k_1, k_2, k_3) = \frac{1}{N_{123}} \sum_{\mathbf{q}_1 \in k_1} \sum_{\mathbf{q}_2 \in k_2} \sum_{\mathbf{q}_3 \in k_3} (2\pi)^3 \delta^{(3)}(\mathbf{q}_1 + \mathbf{q}_2 + \mathbf{q}_3)\delta(\mathbf{q}_1)\delta(\mathbf{q}_2)\delta(\mathbf{q}_3).
\end{equation}
Here $N_{123}$ counts the number of triangle configurations within the bin. The bispectrum is binned with a width of $\Delta k = 3k_f$, where the first bin starts at $k = 1.5 k_f$, where again the smallest scale is $k_{\rm{max}} = 82.5 k_f$, resulting in 2276 triangle bins. 

To calculate the power spectrum and bispectrum for each density field, we use the open-source code BFast\footnote{\href{https://github.com/tsfloss/BFast}{https://github.com/tsfloss/BFast}}, which is GPU accelerated through JAX. Note that the code applies the usual compensation for the mass assignment PCS window function in Fourier space \citep{jing2005correcting}, which was used to construct the density fields. 

Having measured the monopole and quadrupole power spectrum and monopole bispectrum for both pre- and post-reconstructed density fields, which includes 15,000 fiducial simulations and 1500 (including the extra 1000 rotated fields) simulations with varying cosmological parameters as listed in Table \ref{varying parameters}, we have obtained our data products:
\begin{equation}
\mathbf{D} = \left\{P_{pre}^{(0)},P_{pre}^{(2)},B_{pre},P_{post}^{(0)},P_{post}^{(2)},B_{post}\right\}.
\end{equation}

As the halo field is a stochastic tracer of the underlying density field, the reconstructed density field will contain a stochastic component, due to the slightly varying reconstruction quality per halo field. This stochastic component leaves an imprint in the correlation functions, similar to the shot noise contribution to the halo power spectrum, but now scale dependent. We choose not to attempt to model this stochastic contribution and thus do not perform any subtraction of the stochastic component. However, in the following Fisher forecast, the simulation-based covariance matrix does capture the additional variance coming from this implicit stochastic contribution.

\subsection{Covariance}
\begin{figure}[H]
    \centering
    \includegraphics[scale = 0.55]{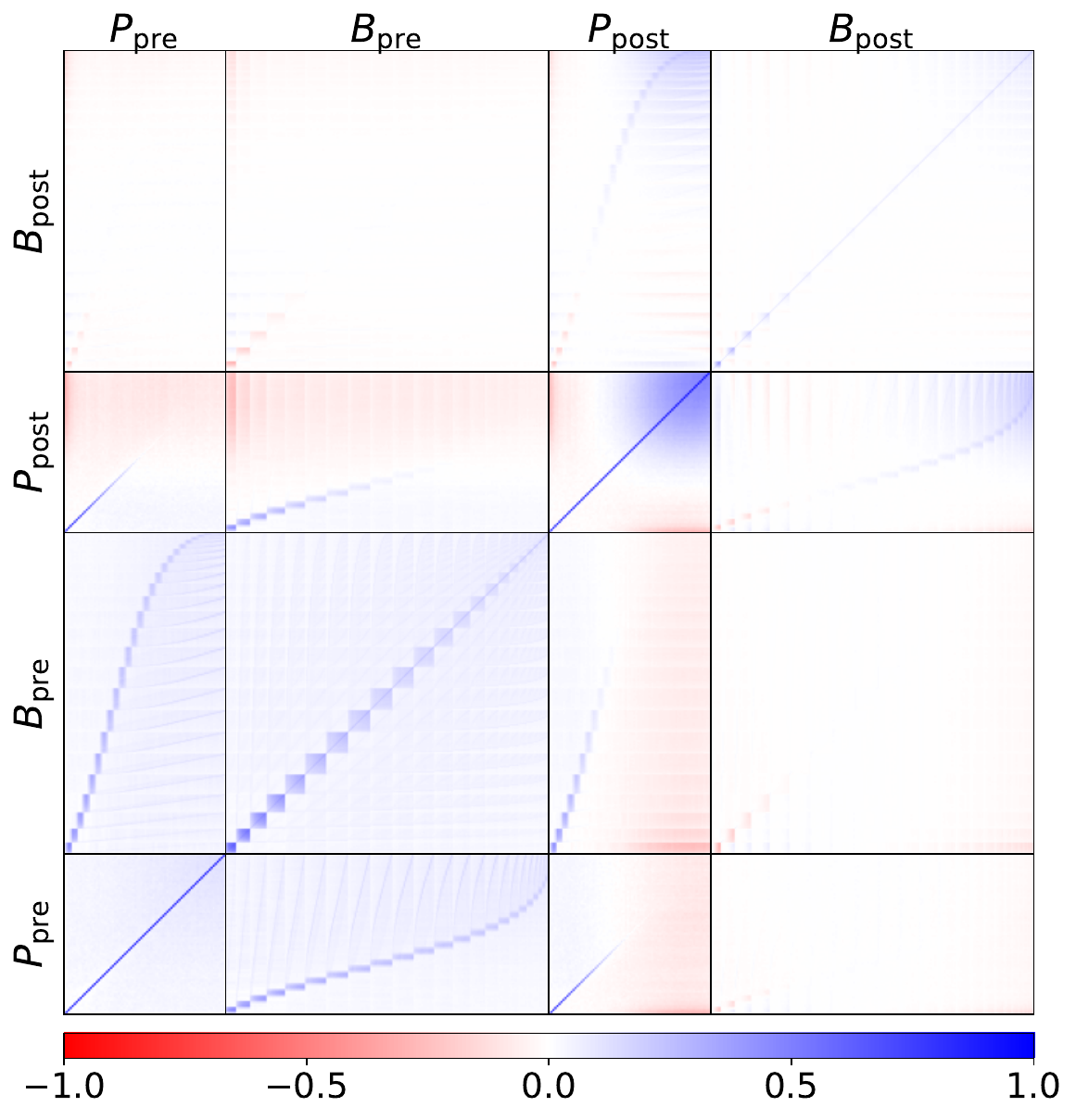}
    \caption{Correlation matrix $r_{ij}$ for the initial and reconstructed power spectrum and bispectrum. Bispectrum triangle configurations are sorted by increasing smallest momentum, creating visual blocks with the same shortest side in $k$-space. Note that we have only plotted the monopole power spectrum for visual clarity, the quadrupole power spectrum results in
    approximately the same correlation matrix as the monopole power spectrum.}
    \label{decoupling of states}
\end{figure}

In order to assess the amount of cosmological information that can be extracted from a set of data, we build the covariance matrix $C$ of the data vector $\mathbf{D}$:
\begin{equation}
\label{covariance_matrix}
    C_{ij} = \frac{1}{N-1}\sum^{N}_{n} (\mathbf{D}_n-\Bar{\mathbf{D}})_i(\mathbf{D}_n-\Bar{\mathbf{D}})_j.
\end{equation}
Here $N$ is the number of simulations and $\mathbf{\Bar{D}}$ is the mean of the data over all $N$ simulations. Subscripts i,j are for indexing the bins in the data vector $\mathbf{D}$. To illustrate the correlation between different values of the power spectrum and bispectrum, we use a correlation matrix:
\begin{equation}
r_{ij} = \frac{C_{ij}}{\sqrt{C_{ii}C_{jj}}}.
\end{equation}
Before reconstruction, modes are coupled, leading to significant non-Gaussian covariance in the measurements. Reconstruction reduces this correlation, as evidenced by the decreased correlation among modes post-reconstruction, demonstrated by the correlation matrix in Figure \ref{decoupling of states}. This reduction in correlation implies the presence of more unique information in the power spectrum and bispectrum of the reconstructed density field.

\subsection{Fisher analysis}

\subsubsection{The Standard Fisher Estimator}

The Fisher matrix formalism \citep{Fisher1922mathematical} provides a framework for estimating parameter uncertainties. Assuming Gaussian likelihood, which is accurate enough for our purposes \citep{scoccimarro2000bispectrum, carron2013assumption}, the Fisher matrix for cosmological parameters $\theta_a$ and data vector $\mathbf{\Bar{D}}$ can be expressed as:
\begin{equation}
\label{Fisher matrix data}
    F^{\theta}_{ab} = \frac{\partial  \mathbf{\bar{D}}_i}{\partial \theta_a} 
     \left(C^{-1}\right)_{ij} \frac{\partial  \mathbf{\bar{D}}_j}{\partial \theta_b}.
\end{equation}
$\mathbf{\Bar{D}}$ is the simulation-averaged data vector, and $C^{-1}$ is the precision matrix derived from the inverse of the covariance matrix from Equation \eqref{covariance_matrix}, adjusted using the Hartlap factor \citep{hartlap2007your, anderson2003introduction}:
\begin{equation}
\label{hartlap}    
    C^{-1} \equiv \frac{N_{\rm{sims}} - N_{\rm{bins}} - 2}{N_{\rm{sims}} - 1} C^{-1} \quad \text{for} \quad N_{\rm{bins}} < N_{\rm{sims}} - 2.
\end{equation}
$N_{\rm{sims}}$ and $N_{\rm{bins}}$ denote the number of simulations and data points (power spectrum and bispectrum bins), respectively. We compute parameter derivatives using finite central differencing:
\begin{equation}
\label{central difference}   
\frac{\partial  \mathbf{\bar{D}}}{\partial \theta_a} = \frac{ \mathbf{\bar{D}}_{\theta_a^{\text{fid}} + \delta\theta_a} -  \mathbf{\bar{D}}_{\theta_a^{\text{fid}} - \delta\theta_a}}{2 \delta\theta_a},
\end{equation}
where $\mathbf{\bar{D}}_{\theta^{\text{fid}}_{a} \pm \delta\theta_{a}}$ represents the mean measured data from simulations with perturbed parameters.

\subsubsection{Compressed Fisher Estimator}
The standard Fisher estimator can lead to overestimation due to Monte Carlo noise, particularly in derivatives. In Ref.~\cite{coulton2023estimate}, Coulton \& Wandelt construct an alternative Fisher estimator, based on a compressed statistic, that is oppositely biased to the standard Fisher estimator and can be combined to reduce the noise and accelerate convergence with the number of simulations. We follow the work of Coulton \& Wandelt and refer the reader to that work for more details.\footnote{We employ a slightly different notation from that of Ref.~\cite{coulton2023estimate}, where we strictly reserve subscripts $i,j$ for indexing the bins in the data vector $\mathbf{D}$, and subscripts $a,b,c,d$ for indexing parameters $\theta$ }. An optimal compression, for the parameters from data vector $\mathbf{D}$, would require knowledge of precisely the ingredients of the standard Fisher estimator. Instead, we use a subset $\alpha$ number of simulations to construct the (suboptimally) compressed statistic. For a Gaussian likelihood with parameter independent covariance this is given by \citep{tegmark1997karhunen, heavens2000massive, coulton2023estimate}:
\begin{equation}
\label{compressed statistic}
    t_a =  \frac{\partial  \mathbf{\Bar{D}}_i}{\partial \theta_a} (C^{\alpha})^{-1}_{ij} \left(\mathbf{D}_j - \mathbf{\Bar{D}}^{\alpha }_j\right),
\end{equation}
where $C^{\alpha}$ is the covariance matrix in Equation \eqref{covariance_matrix} constructed from the subset $\alpha$ of simulations, and $\bar{\mathbf{D}}^\alpha$ the mean over the subset $\alpha$. In this compressed space, the Fisher information is:
\begin{equation}
\label{compressed Fisher estimator}
F^{tt}_{ab} = \frac{\partial \mu^t_{c}}{\partial \theta_a} (\Sigma^{tt})^{-1}_{cd} \frac{\partial \mu^t_{d}}{\partial \theta_b}.
\end{equation}
Here $\mu^t_c$ is the mean of the compressed statistic, $\Sigma^{tt}$ is its covariance:
\begin{equation}
    \Sigma^{tt}_{cd} = \frac{\partial  \mathbf{\Bar{D}}^{\alpha}_{i}}{\partial \theta_c} \cdot (C^{\alpha})^{-1}_{ij} \cdot \frac{\partial  \mathbf{\Bar{D}}^{\alpha}_{j}}{\partial \theta_d}.
\end{equation}
The remaining subset of simulations, $\beta$, is used to compute the derivatives to use in Equation (\ref{compressed Fisher estimator}):
\begin{equation}
\label{compressed mean}
\frac{\partial \mu^t_c}{\partial \theta_a} =  \frac{\partial  \mathbf{\Bar{D}}^{\alpha}_i}{\partial \theta_c} \cdot (C^{\alpha})^{-1}_{ij} \cdot \frac{\partial  \mathbf{\Bar{D}}^{\beta}_j}{\partial \theta_a}.
\end{equation}
To avoid bias in the compressed Fisher estimator due to reusing simulations, we split the simulations into two parts: 75\% for estimating the compression and 25\% for computing the quantities for compressed Fisher information. 

\subsubsection{Combined Fisher Estimator}
\label{combined Fisher estimator}
After calculating the standard Fisher estimator, which tends to be biased high, and the compressed Fisher estimate, which tends to be biased low, we can combine these two estimators to compute an unbiased estimator. To create a combined estimator, we use the geometric mean \citep{bhatia2009positive}:
\begin{equation}
F^{\text{Combined}}_{IJ} = G(F^{\text{Standard}}_{IJ}, F^{\text{comp}}_{IJ}),
\end{equation}
with $G(A,B)$ being:
\begin{equation}
G(A, B) = A^{1/2} \left( A^{-1/2} B A^{-1/2} \right)^{1/2} A^{1/2}.
\end{equation}
After obtaining one estimate of the combined Fisher information, we can generate a second estimate by reassigning the simulations between the compression and Fisher estimation steps. This process can be repeated multiple times, generating a set of partially correlated estimates that, when averaged, reduces the estimator variance \citep{coulton2023estimate}. For our purposes, we shuffled the simulations eight times to ensure that our measurements were reliable. The code for calculating the combined Fisher estimator is publicly available
\footnote{\href{https://github.com/wcoulton/CompressedFisher}{https://github.com/wcoulton/CompressedFisher}}. The combined Fisher estimator ensures that we avoid Monte Carlo uncertainties arising from the finite number of simulations, particularly in the derivatives.

\subsection{Marginalization}
For the Fisher matrices, the marginalized error is given by the Cramer-Rao bound \citep{Rao1945, Cramer1946}, under the assumption of a Gaussian likelihood:
\begin{equation}
\label{Cramer-Rao}
\sigma_{\theta_a} = \sqrt{(F^{-1})_{aa}}.
\end{equation}

\begin{figure}
    \centering
    \includegraphics[scale = 0.75]{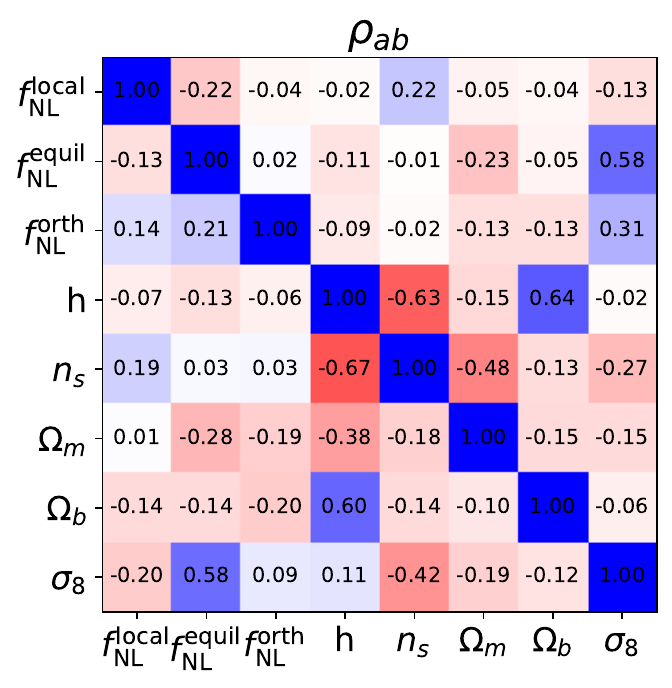}
    \caption{Correlation coefficients $\rho_{ij} \in [-1, 1]$ between the different cosmological parameters when using the data products \textcolor{softgreen}{$(P + B)_{\text{pre}}$} (lower triangular) and \textcolor{softblue}{$(P + B)_{\text{pre}} + (P + B)_{\text{post}}$} (upper triangular).}
    \label{degenaracy visualized}
\end{figure}

In the Fisher analysis, we can also quantify the degeneracy between parameters by their correlation coefficient:
\begin{equation}
\label{degenaracy}
    \rho_{ab} = \frac{(F^{-1})_{ab}}{\sqrt{\sigma_{a}^{2} \sigma_{b}^{2}}}.
\end{equation}
Figure \ref{degenaracy visualized} shows a reduced degeneracy between the pnG parameters and between the pnG parameters and the cosmological parameters. This implies that reconstruction helps in isolating the effects of the pnG parameters.

Furthermore, the Fisher matrix can be used to construct an unbiased, minimal-variance marginalized estimator that estimates the parameter $\theta^{a}$ from the observed data $\mathbf{D}_{\text{obs}}$ given a fiducial model 
$\mathbf{\bar{D}}_{\text{fid}}$:
\begin{equation}
\label{bias of estimator}
\hat{\theta}^{a} - \theta_{\text{fid}}^{a} = \sum_{b} (F^{-1})_{ab} \frac{\partial \mathbf{\bar{D}}}{\partial \theta^{b}} \cdot C^{-1} \cdot (\mathbf{D}_{\text{obs}} - \mathbf{\bar{D}}_{\text{fid}}). 
\end{equation}
Since the network is trained only on simulations following the fiducial cosmological model, reconstructions derived from simulations with different parameters may have inherent biases. This bias could potentially be propagated through the process when these reconstructions are used to compute derivatives, shown in Equation \eqref{central difference}. Despite this, the marginalization process is designed to incorporate and account for such biases. Therefore, it is expected that estimates of the parameters will remain unbiased and achieve minimal variance, even if the cosmology of the observed data deviates from the fiducial model.
\subsection{Parameter Constraints}
\label{Results section}

Table \ref{table all Fisher} displays the (un)marginalized errors on the pnG parameters, comparing the \textcolor{softgreen}{pre}, \textcolor{softred}{post}, and \textcolor{softblue}{(pre+post)} data points across power spectra (P), bispectra (B), and joint power spectra and bispectra (P+B). 
\begin{table}
\centering
\begin{tabular}{lccc}
\hline
&  $f_{\rm{NL}}^{\rm{local}}$ & $f_{\rm{NL}}^{\rm{equil}}$ & $f_{\rm{NL}}^{\rm{orth}}$\\ 
\hline

\textcolor{softgreen}{$P_{\text{pre}}$} & \textcolor{softgreen}{199.26 (74.90)} & \textcolor{softgreen}{853.81 (217.63)} & \textcolor{softgreen}{253.91 (75.72)} \\
\textcolor{softred}{$P_{\text{post}}$} & \textcolor{softred}{184.13 (91.38)} & \textcolor{softred}{368.90 (86.65)} & \textcolor{softred}{137.23 (83,03)} \\
\textcolor{softblue}{$P_{\text{pre}} + P_{\text{post}}$} & \textcolor{softblue}{128.41 (69.81)} & \textcolor{softblue}{227.51 (80.62)} & \textcolor{softblue}{76.31 (53.26)} \\
\hline
\textcolor{softgreen}{$B_{\text{pre}}$} & \textcolor{softgreen}{135.90 (76.31)} & \textcolor{softgreen}{188.33 (217.78)} & \textcolor{softgreen}{64.15 (75.49)} \\
\textcolor{softred}{$B_{\text{post}}$} & \textcolor{softred}{100.40 (69.04)} & \textcolor{softred}{100.40 (74.35)} & \textcolor{softred}{37.58 (31.79)} \\
\textcolor{softblue}{$B_{\text{pre}} + B_{\text{post}}$} & \textcolor{softblue}{80.56 (62.00)} & \textcolor{softblue}{90.29 (66.94)} & \textcolor{softblue}{34.39 (29.61)} \\
\hline
\textcolor{softgreen}{$(P + B)_{\text{pre}}$} & \textcolor{softgreen}{86.10 (55.32)} & \textcolor{softgreen}{159.39 (111.19)} & \textcolor{softgreen}{47.25 (40.88)} \\
\textcolor{softred}{$(P + B)_{\text{post}}$} & \textcolor{softred}{86.17 (53.78)} & \textcolor{softred}{90.83 (64.88)} & \textcolor{softred}{34.12 (29.26)} \\
\textcolor{softblue}{$(P + B)_{\text{pre}} + (P + B)_{\text{post}}$} & \textcolor{softblue}{64.95 (46.37)} & \textcolor{softblue}{84.61 (61.97)} & \textcolor{softblue}{30.13 (26.56)} \\
\hline
\end{tabular}

\caption{(Un)marginalized errors on $f^{\rm{shape}}_{\rm{NL}}$ for all options of statistics estimated with the combined Fisher estimator from Section \ref{combined Fisher estimator}.}
\label{table all Fisher}
\end{table}
The decoupling of modes, the mitigation of redshift-space distortions, and focusing on pnG parameters the reduction of parameter degeneracy are the main factors contributing to this improvement. The improvement factors, quantified by the ratio of these marginalized errors, are presented in Table \ref{improv factor}.
\begin{table}
    \centering
    \begin{tabular}{|c|c|c|c|c|c|c|c|}
    \hline
        $f_{\rm{NL}}^{\rm{local}}$ & $f_{\rm{NL}}^{\rm{equil}}$ & $f_{\rm{NL}}^{\rm{orth}}$ & $h$ & $n_s$ & $\Omega_m$ & $\Omega_b$ & $\sigma_8$\\
        \hline
        1.33 & 1.88 & 1.57 & 1.75 & 1.44 & 1.55 & 1.76 & 1.51  \\
        \hline
    \end{tabular}
    \caption{Improvement factor of parameter constraints when using the full data product
\textcolor{softblue}{$(P + B)_{\rm{pre}}$ + $(P + B)_{\rm{post}}$} as compared to just \textcolor{softgreen}{$(P + B)_{\rm{pre}}$}.}
    \label{improv factor}
\end{table}
An exception in the improvement pattern is the local pnG parameter $f_{\rm{NL}}^{\rm{local}}$, which shows a relatively lower improvement factor compared to the equilateral $f_{\rm{NL}}^{\rm{equil}}$ and orthogonal $f_{\rm{NL}}^{\rm{ortho}}$ pnG parameters. This difference can be attributed to the scale-dependent bias present in the \textcolor{softgreen}{pre}-reconstruction dark matter halo field, a phenomenon well documented in previous studies \citep{Wagner2012, Scoccimarro2012, Sefusatti2012, coulton2022quijote}. Figure \ref{scale dependent bias plot} in the Appendix, confirms the presence of this bias, which enhances the informative power of the power spectrum for $f_{\rm{NL}}^{\rm{local}}$.

\newpage

\begin{figure}
    \centering
    \includegraphics[scale = 0.61]{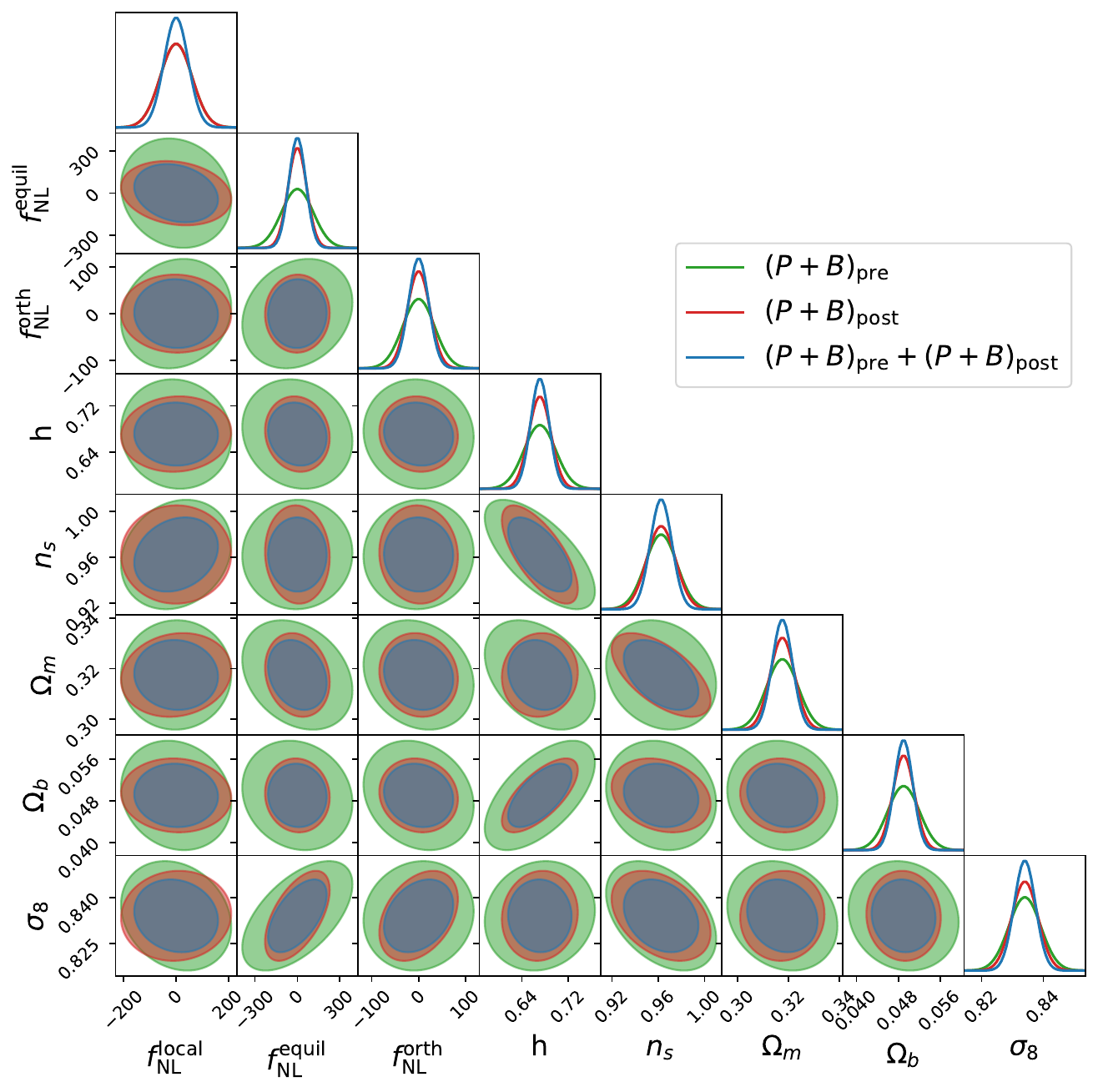}
    \caption{Comparison of the constraining power for the \textcolor{softgreen}{pre}, \textcolor{softred}{post}, and \textcolor{softblue}{pre+post} measured power spectrum and bispectrum, jointly fitted for all 8 parameters. The coloured areas indicate the $2\sigma$ constraints, highlighting the sensitivities and different degeneracy directions. The diagonal plots represent the marginalized probability distributions for each parameter, while the off-diagonal plots illustrate the confidence intervals for the parameter pairs.}
    \label{marginalized errors 8 params}
\end{figure}

Figure \ref{marginalized errors 8 params} displays the resulting confidence intervals and the marginalized combined Fisher errors for the eight parameters (five cosmological parameters and three pnG parameters) when fitting them simultaneously, using both power spectrum and bispectrum data.

\section{Conclusions}
\label{chapter 9}
We have presented a neural network-based approach to reconstructing the linear density field from late-time dark matter halo data. Our findings demonstrate that the reconstruction quality for a $1 \text{Gpc}^3$ volume with average number density of halos $\bar{n} = 4 \times 10^{-4}$ (h/Mpc)$^3$, as measured by the cross-correlation with the linear dark matter density field, achieves approximately 44\% accuracy for scales $k\leq 0.4 \text{h}/\text{Mpc}$. This is a strong improvement over the cross-correlation of the unreconstructed halo fields with the initial dark matter density field. We furthermore find strong evidence that our reconstruction is optimal for this data.

By performing the reconstruction and a subsequent Fisher analysis on a large part of the Quijote simulation suite, we show that this reconstruction leads to a significant overall improvement in cosmological parameter constraints. Focussing on primordial non-Gaussianity, we find that the marginalized errors on pnG from a joint analysis of the power spectrum (monopole and quadrupole) and bispectrum are improved by factors of 1.33, 1.88, and 1.57 for local, equilateral, and orthogonal pnG. This improvement is obtained, because the reconstruction method mitigates redshift-space distortions, decouples modes, and decreases degeneracies among cosmological parameters. Since shot noise and non-Gaussian covariance cause saturation of parameter constraints from the dark matter halo field power spectrum and bispectrum \citep{coulton2022quijote}, such improvements are not easily achieved by simply including more modes in the halo field analysis. The improvement is significantly lower compared to the dark matter case in \cite{Thomas}, since reconstruction is limited by shot noise in the dark matter halo fields. Nevertheless, our reconstruction method could potentially be applied to complement existing observational data analysis methods, leading to improvements in cosmological parameter constraints. Improvements that would otherwise require an increase in number density \citep{DAmico:2022gki}, or a larger survey volume \citep{Cabass:2022epm}, both of which are costly and observationally challenging compared to our reconstruction method.

A generic limitation of the type of reconstruction performed in this work is the GPU memory constraint that restricted our grid resolution to $256^3$ cells, limiting the amount of modes input to the network to $k \leq 1.4$ h/Mpc. Although we find evidence that in our case (i.e. resolution and number density) the reconstruction is already optimal, for higher number densities this limited resolution could limit the accuracy of reconstruction, especially on smaller scales. However, on small scales, shell crossing of dark matter streams, and virialization of bound structures puts a fundamental limit on the reconstructability of linear modes, implying diminishing returns from including additional modes. Further research is required to determine this limit. Also, the effect of the stochastic component after reconstruction is not well understood but could lead to tighter parameter constraints when accurate noise subtraction is applied \citep{Chan:2016ehg}. The challenge in determining the exact stochastic contribution is in part due to the low interpretability of the neural network. 
Moreover, while \cite{scoccimarro2000bispectrum} justified the assumption of a Gaussian likelihood in our statistics for the Fisher analysis, \cite{park2023quantification} demonstrated that this assumption can lead to artificially tight bounds on parameter constraints when the statistics are not truly Gaussian. Also, the reconstruction is performed with a fixed bias term $b_1$, which means we could not marginalize over this term in the analysis. Another concern is that in simulations, the exact positions of dark matter halo tracers are known in redshift space, which may not be achievable in real observational data.

Future research could focus on applying this reconstruction method to real tracers such as galaxies or atomic spectral lines instead of dark matter halo fields. Although in this study we have used the halos as a proxy for such tracers, for practical application on observational data, this method should be further developed on even more realistic tracers. It would also be worth exploring different inputs beyond just the halo number density. For instance, incorporating velocity or void information could potentially make the input more informative for the U-Net in reconstructing the initial density distribution \cite{Bayer:2022vid}. Studying the impact of these additions on reconstruction accuracy would be interesting and could lead to better results depending on the information available in actual observational data.

This work was motivated by a previous study on the dark matter field. We have demonstrated that reconstructing the initial conditions using a U-Net architecture neural network model on $z=0$ simulated dark matter halo fields can provide non-trivial improvements on cosmological parameter constraints. Our analysis shows that the U-Net can, to a certain degree, reverse the non-linear collapse of matter, and decoupling modes. The resulting field is more linear and Gaussian, and its power spectrum and bispectrum contain additional information on cosmological parameters. We have identified several avenues for further investigation, with the ultimate goal to explore the potential of these reconstructions on real data. 

\acknowledgments

The authors like to thank Will Coulton and Florian List for comments on the manuscript. We thank the Quijote team for making the simulation suite available \citep{villaescusa2020quijote}. We thank the Center for Information Technology of the University of Groningen for providing access to the Hábrók high-performance computing clusters. We thank the developers of NumPy, JAX, Matplotlib, Optax for providing the open-source tools used in this research.

\bibliographystyle{JHEP}
\bibliography{bibliography.bib}

\newpage
\appendix

\section{Modifying shot noise}
\label{modifying shot noise}
Since shot noise is the main cause of saturation of cosmological parameter constraints \citep{coulton2022quijote}, it is worth exploring how varying levels of shot noise in the input fields affect the accuracy of reconstruction, and thereby parameter constraints. We have examined two scenarios: increasing the halo mass threshold, which decreases the halo number density and thus increases shot noise by definition, and applying halo mass weighting when constructing the halo density fields, as this has been shown to effectively reduce shot noise \citep{hamaus2010minimizing}. The simulated halo catalogues provide detailed mass information, offering more insights into the matter distribution. By comparing reconstructions with and without mass information, we assess the impact of this additional information on the quality of the reconstruction.

When generating halo density fields from the halo catalogue, each halo's mass is given, and determined by the number of dark matter particles it contains. For the main analysis in this paper, we used the lowest halo mass threshold $M_{\rm{min}} \geq 1.32\cdot 10^{13} \: M_{\odot}/\text{h}$ (20 dark matter particles) from the Quijote simulation suite, which results in an average number density of approximately $n \geq 4 \cdot 10^{-4} \text{ h}^3/\text{Mpc}^3$. Raising the threshold decreases the number density, which inherently amplifies shot noise. This adjustment enables us to evaluate the performance of the reconstruction method under conditions of increased shot noise. 

Mass information available from observational data could improve reconstruction accuracy compared to the basic setup. To investigate this, we classified halos into mass bins, with higher bin numbers indicating more massive halos. Increasing the number of bins provides more detailed mass information. Next, we can incorporate this mass information in our 3D density fields with the mass weighting using PCS, reducing the shot noise \citep{hamaus2010minimizing} in the fields, allowing us to assess the impact of varying degrees of detail in mass weighting on reconstruction accuracy.

\begin{figure}[h]
    \centering
    % First subfigure
    \begin{subfigure}[b]{0.47\textwidth}
        \centering
        \includegraphics[width=\textwidth]{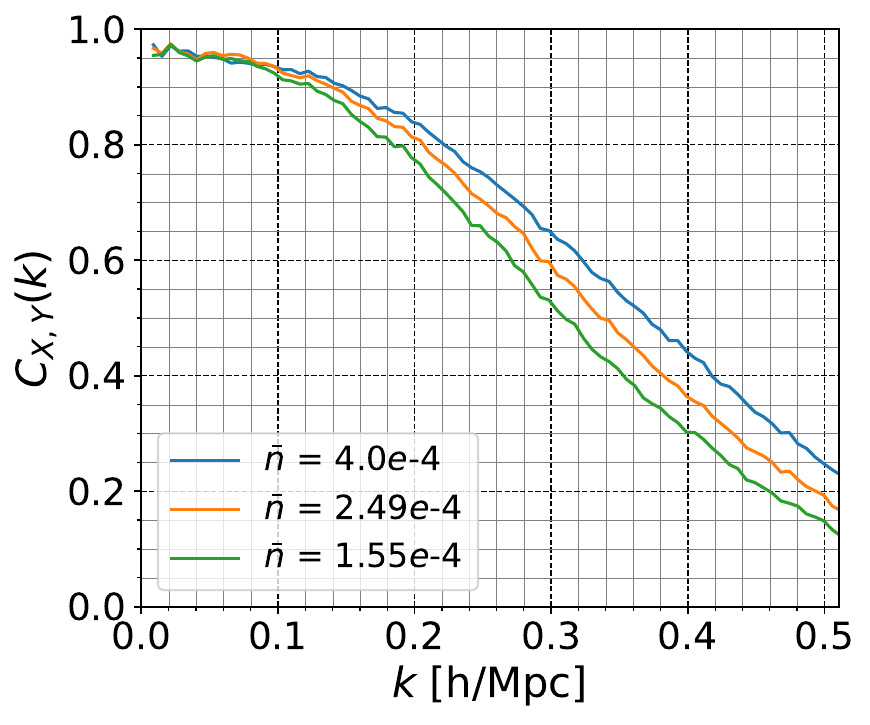}
        \caption{}
        \label{number density cross corr}
    \end{subfigure}
    \hfill
    % Second subfigure
    \begin{subfigure}[b]{0.47\textwidth}
        \centering
        \includegraphics[width=\textwidth]{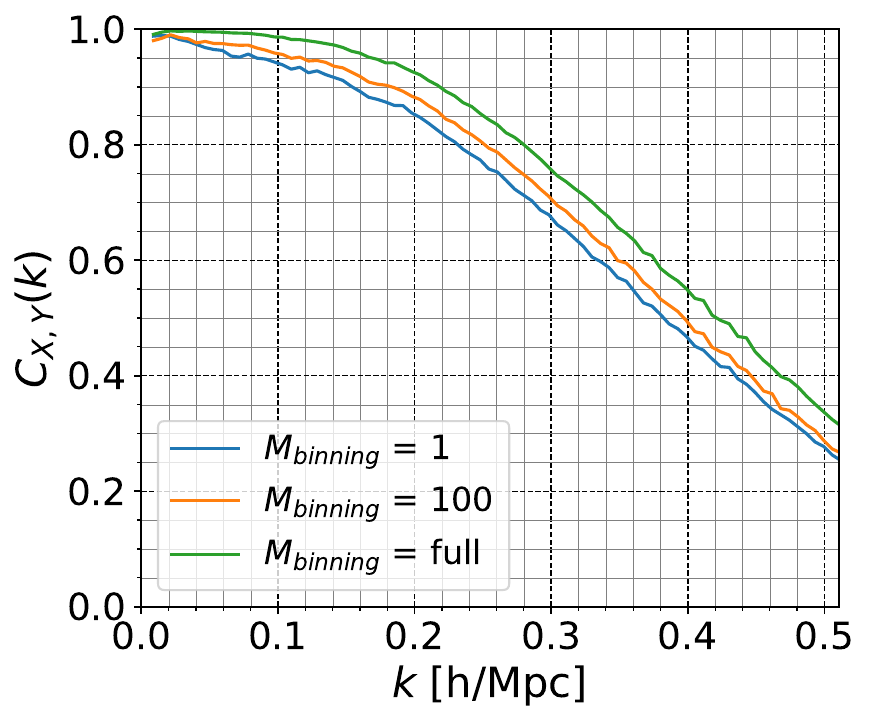}
        \caption{}
        \label{mass information cross corr}
    \end{subfigure}
    % Overall figure caption
    \caption{Both sub-figures display the cross-correlation of reconstructed density fields with varying shot noise. (a) shows the impact of varying halo number density on reconstruction quality, and (b) shows the effect of different mass weightings accuracy in the density fields on reconstruction quality. Here, full means that the model has the exact mass information of each dark matter halo.\protect\footnotemark}
    \label{fig:combined}
\end{figure}

First, we constructed density fields with an increased minimum halo mass threshold. These density fields were used to train the U-Net. To evaluate the changes in reconstruction quality by varying the number density of the initial field, we compared the cross-correlation with the initial density field. The results are shown in Figure \ref{number density cross corr}. As expected, the quality of reconstruction decreases with decreasing number density. On large scales up to $k=0.1\text{ h}/\text{Mpc}$, the reconstruction quality is approximately the same for different number densities. However, for lower number densities, the reconstruction clearly deteriorated more rapidly at smaller scales.

Second, we explore how our model changes with different levels of detail in mass information. Until now, we have only used the number density of dark matter halos to train the model. However, in galaxy surveys, an estimate of galaxy masses could provide additional information on the distribution of matter, effectively reducing shot noise \cite{hamaus2010minimizing}. As an analogue, the Quijote halo catalogue provides specific information about the mass of each halo. This additional mass information could potentially enhance the model's ability to reconstruct the initial conditions. To investigate the extent of this improvement, we train different models with varying levels of mass information. By again comparing the cross-correlations for these different cases, we study how our reconstruction method benefits from having more detailed mass distribution information (reduced shot noise). Figure \ref{mass information cross corr} illustrates that incorporating additional mass information into the input density fields improves the quality of reconstruction. This improvement arises because the model gains more insight into the underlying dark matter field compared to scenarios without mass information.

\footnotetext{\par
This differs from the underlying dark matter field because the halo threshold is set to 20 dark matter particles. Consequently, regions below this threshold are devoid of dark matter halos.}

\section{Convergence of Fisher Analysis}
\label{convergence section}
\begin{figure}[h]
\centering
% First row with two subfigures
\begin{subfigure}{0.45\textwidth}
    \centering
    \includegraphics[width=\textwidth]{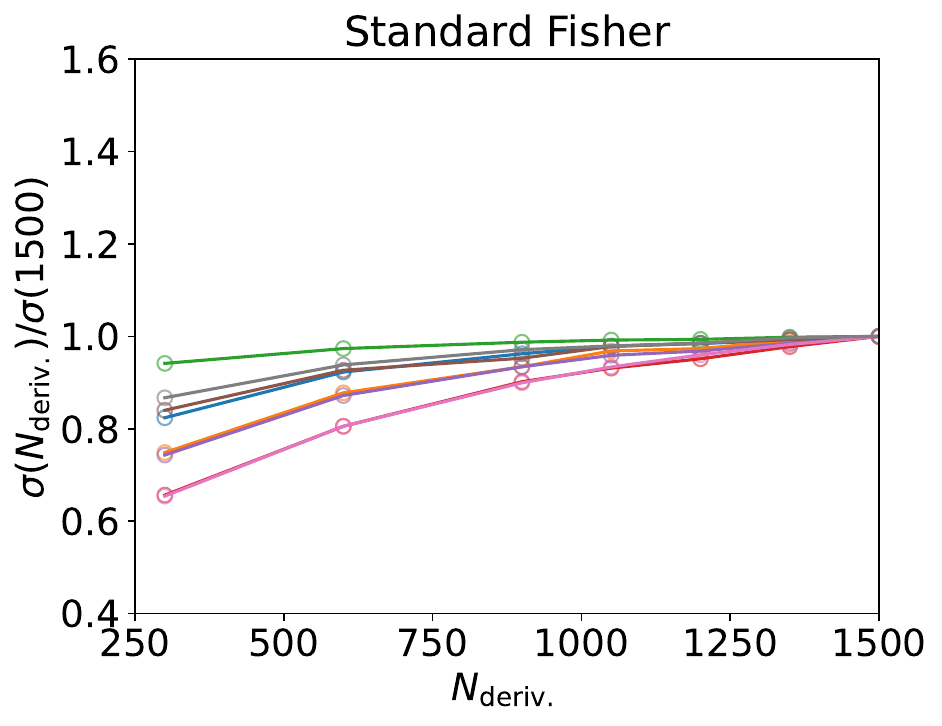}
    \caption{Standard Fisher}
    \label{standard_Fisher}
\end{subfigure}
\hfill % adds horizontal space between the figures
\begin{subfigure}{0.45\textwidth}
    \centering
    \includegraphics[width=\textwidth]{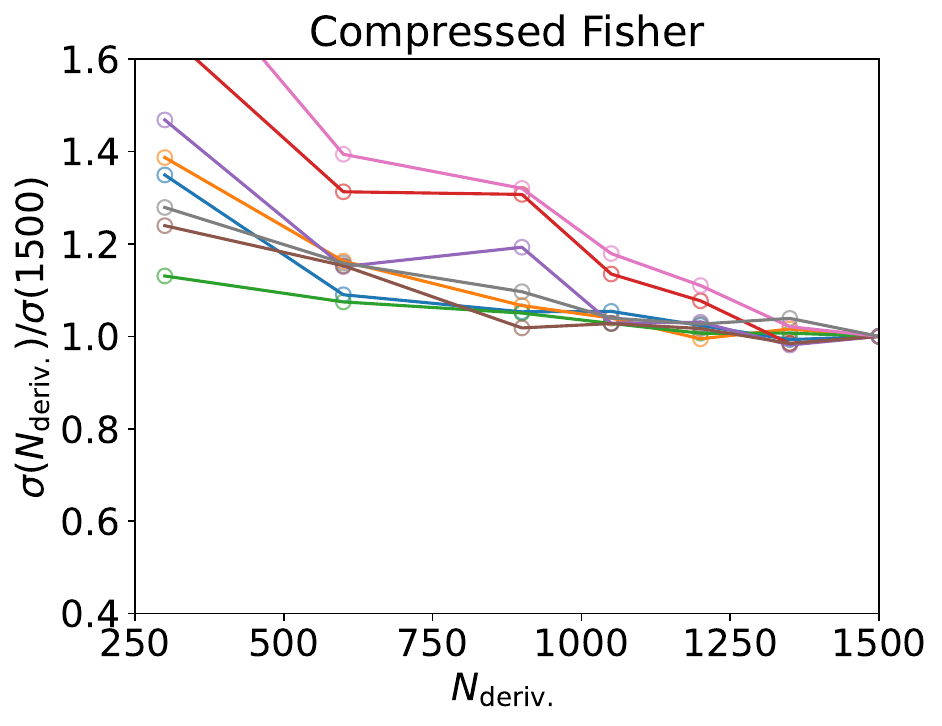}
    \caption{Compressed Fisher}
    \label{compressed_Fisher}
\end{subfigure}

\vspace{0.01cm} % adds vertical space between the rows

% Second row with one subfigure
\begin{subfigure}{0.6\textwidth}
    \centering
    \includegraphics[width=\textwidth]{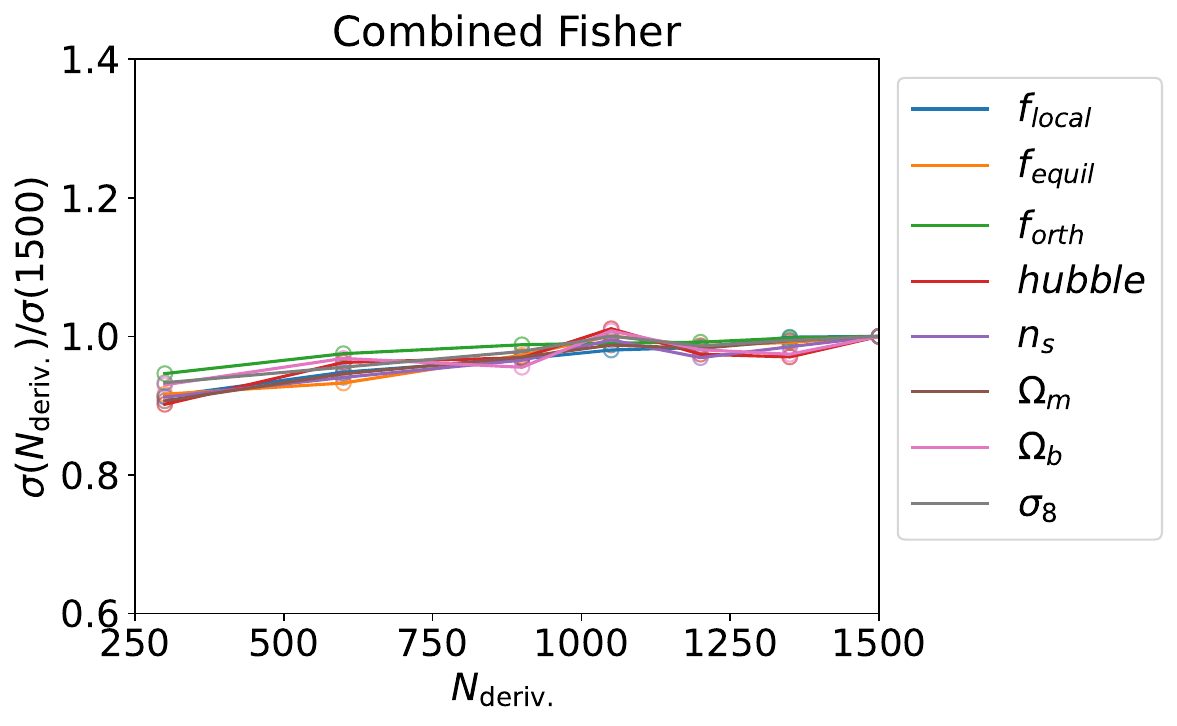}
    \caption{Combined Fisher}
    \label{combined_Fisher}
\end{subfigure}

\caption{Convergence of Fisher matrix estimates for \textcolor{softblue}{($P+B)_{pre}+(P+B)_{post}$}: (a) Standard Fisher, (b) Compressed Fisher, and (c) Combined Fisher. The combined estimator balances the biases present in the standard and compressed Fisher matrices, resulting in unbiased estimates.}
\label{Fisher_convergence}
\end{figure}

In Section \ref{combined Fisher estimator} we introduced the usage of an unbiased combined Fisher matrix that takes the geometric mean from a biased high standard Fisher matrix and a biased low compressed Fisher matrix, in order to overcome the Monte Carlo uncertainties arising from the limited number of simulations to compute the numerical derivatives \citep{coulton2023estimate}. Figure \ref{standard_Fisher} demonstrates the behaviour of the standard Fisher matrix estimate as the number of derivatives, $N_{\rm{deriv}}$, increases. As expected, the standard Fisher information tends to be biased high, indicated by the upward trend. 

In Figure \ref{compressed_Fisher}, the compressed Fisher matrix estimate shows a different trend. Here, the values tend to decrease as $N_{\rm{deriv}}$ increases, indicating that the compressed Fisher information is biased low. This low bias is due to the reduction of noise in the derivatives of the compressed statistic \citep{coulton2023estimate}.

\begin{figure}[H]
    \centering
    \includegraphics[scale= 0.50]{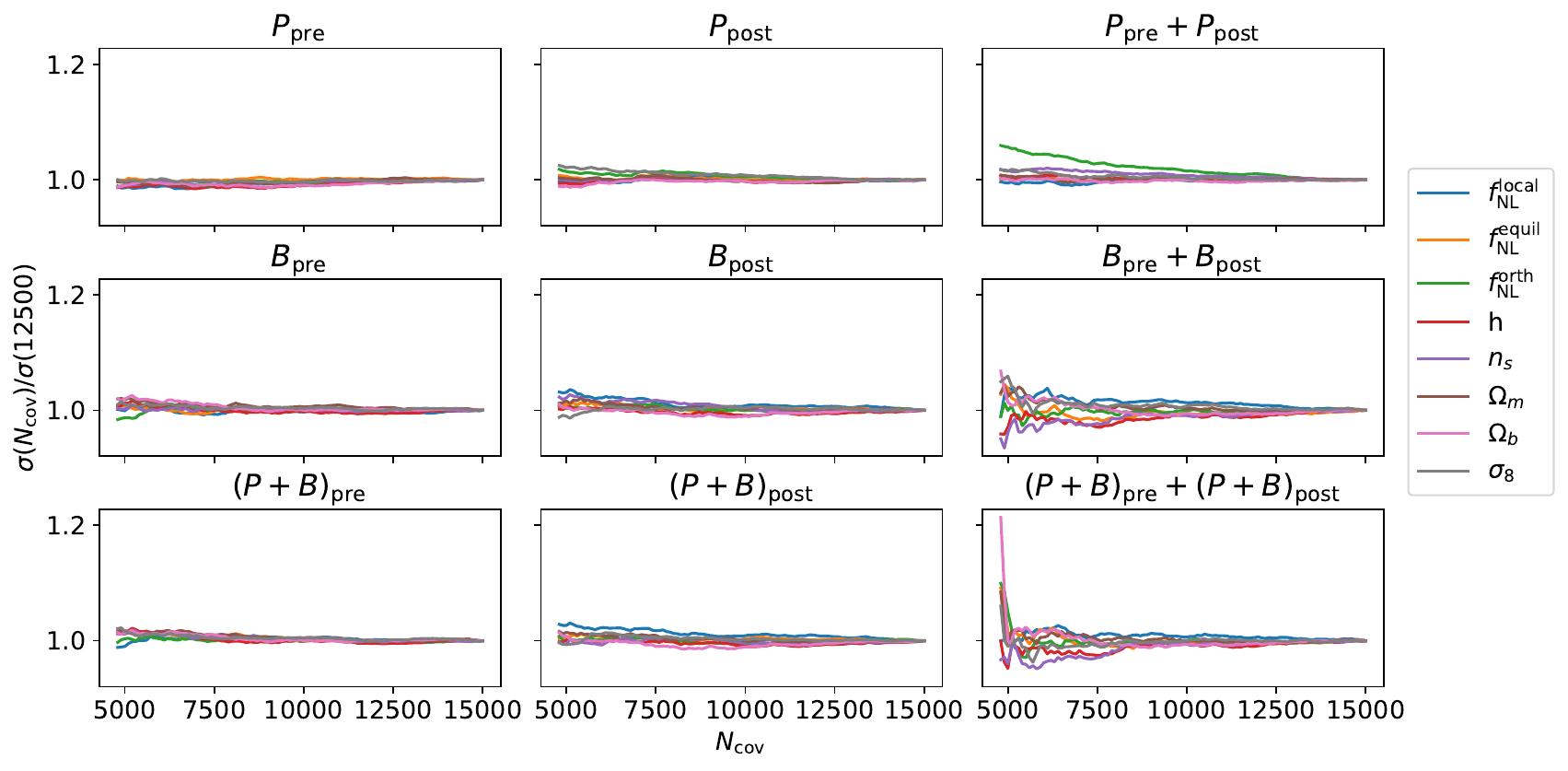}
    \caption{Variation of the marginalized parameter constraints as a function of the number of simulations included in the computation of the covariance matrix. The figure shows that the covariance is converged for all the scenarios analysed in this thesis}
    \label{Covariance covergence}
\end{figure}

Figure \ref{combined_Fisher} presents the combined Fisher matrix estimate, which merges the standard and compressed estimates using the geometric mean approach. This combined estimator is converged, stabilizing around a value close to 1. The convergence observed in the combined Fisher matrix suggests that it balances the high bias of the standard Fisher matrix and the low bias of the compressed Fisher matrix. The results validate the combined estimator as an unbiased method for computing the Fisher matrix, which is also validated from the minimal-variance marginalized estimator in Equation \eqref{bias of estimator}). Furthermore, the convergence of the covariance matrix is demonstrated in Figure \ref{Covariance covergence}.

\newpage

\newpage

\section*{Scale-dependent bias}
\begin{figure}[h]
    \centering
    \includegraphics[width = 0.9\textwidth]{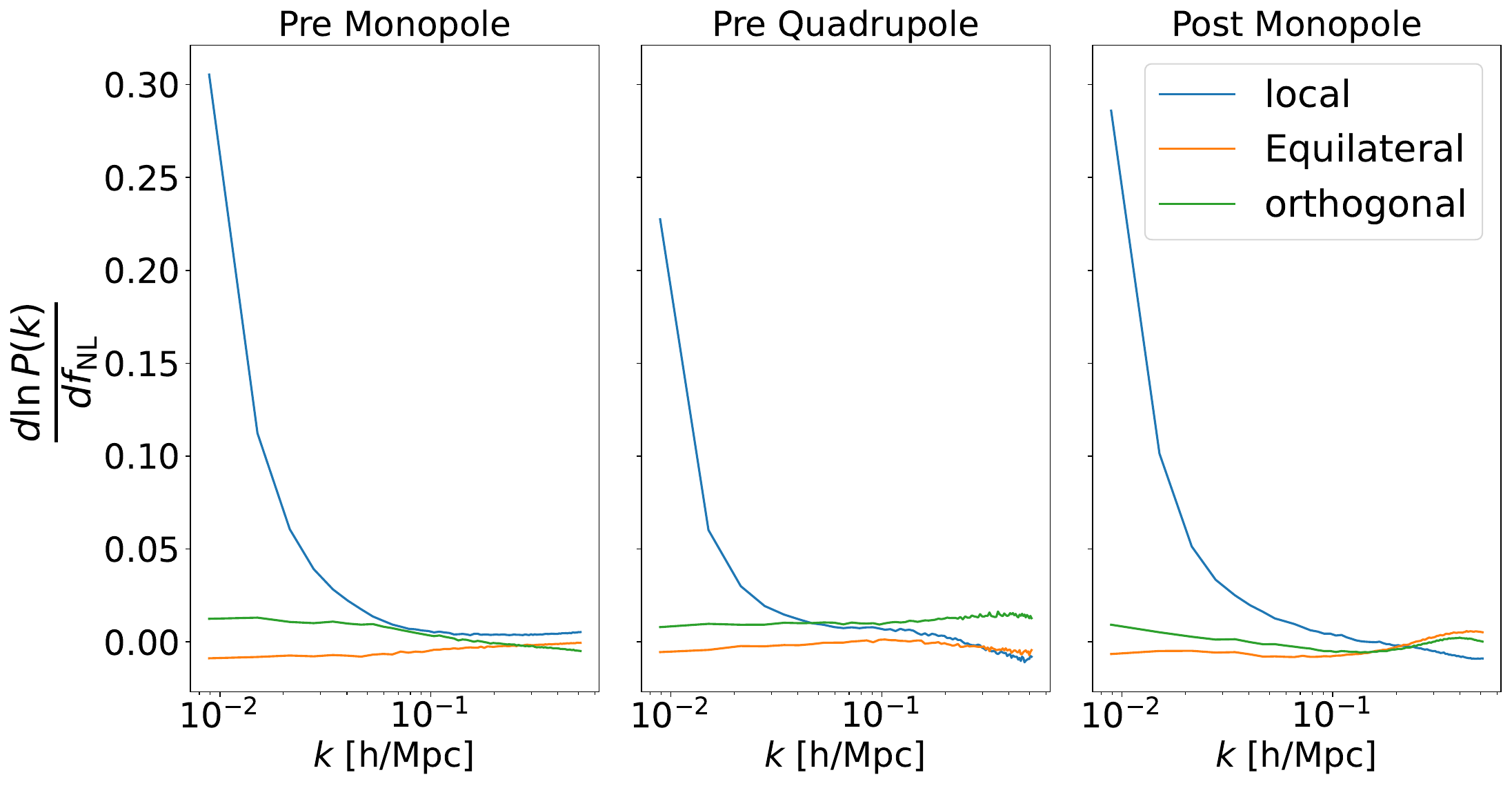}
    \caption{The derivative of power spectra with respect to the amplitudes of the three types of pnG, monopole, and quadrupole plotted for pre-reconstruction and monopole for post-reconstruction. It highlights the scale-dependent bias present in local pnG, which explains the low improvement factor after reconstruction. This is because information from smaller scales is already incorporated into the larger scales in the pre-reconstructed power spectrum.}
    \label{scale dependent bias plot}
\end{figure}

\section*{Reconstruction without gradient field contribution}
\label{Without PT section}\

\begin{figure}[h]
    \centering
    \includegraphics[width=0.5\linewidth]{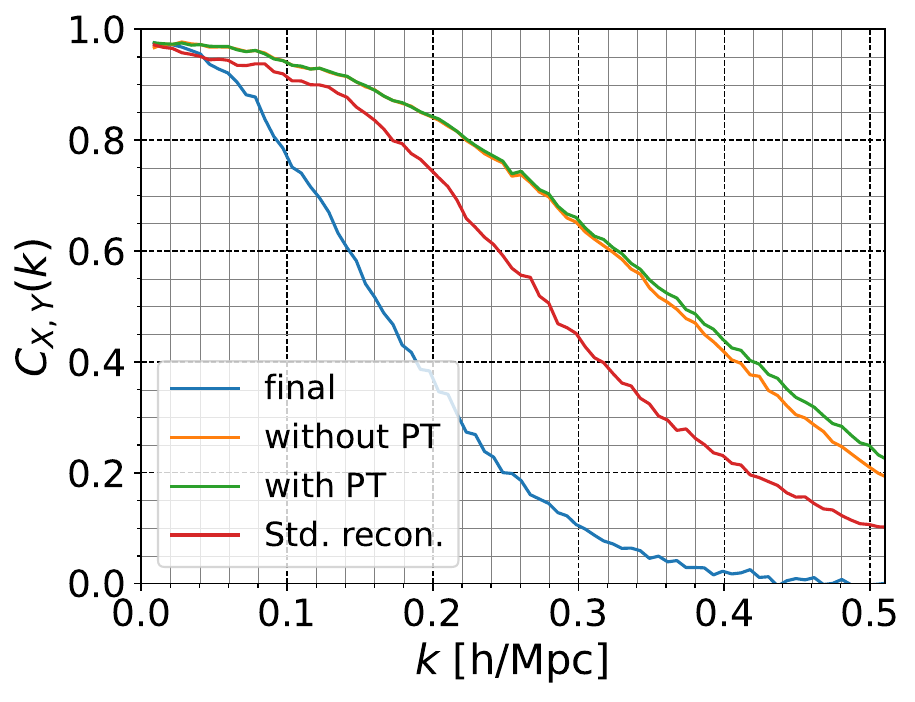}
    \caption{The cross-correlation of density fields at a single realization. The orange line shows the performance of the model without the gradient fields in equations \eqref{velocity field} and \eqref{deriv velocity field}, and is compared to the model that incorporated these gradient fields }
    \label{without PT}
\end{figure}

To showcase the use of inputting the gradient fields in equations \eqref{velocity field} and \eqref{deriv velocity field}, we also trained the model without incorporating the gradient fields. Since these fields do not introduce additional information to the network beyond what is already contained in the density field, they were primarily introduced to improve computational efficiency and convergence. This is supported by Figure \ref{without PT}, which shows that the model without PT achieves nearly the same reconstruction accuracy. However, more training time was required to reach this level of performance. Depending on the complexity of the network, it might still not reach the same accuracy achieved when incorporating the gradient fields.


%\begin{thebibliography}{99}

%\bibitem{a}
%Author, \emph{Title}, \emph{J. Abbrev.} {\bf vol} (year) pg.

%\bibitem{b}
%Author, \emph{Title},
%arxiv:1234.5678.

%\bibitem{c}
%Author, \emph{Title},
%Publisher (year).

% Please avoid comments such as "For a review'', "For some examples",
% "and references therein" or move them in the text. In general,
% please leave only references in the bibliography and move all
% accessory text in footnotes.

% Also, please have only one work for each \bibitem.

%\end{thebibliography}
\end{document}